\documentclass[trackchanges,twocolumn]{aastex701}
\usepackage{amsmath}

\newcommand{\lbol}{L_\mathrm{bol}}
\newcommand{\lbha}{L_\mathrm{H\alpha,broad}}

\begin{document}

\shorttitle{Scaling Relation of LRDs}

\title{
ATLAS. I. \\
A Scaling Relation of LRDs between Broad H$\alpha$ and Bolometric Luminosities:\\
Enhanced Broad H$\alpha$ Emission Relative to Low-$z$ Type 1 AGN


}

\author[orcid=0009-0006-6763-4245]{Hiroto Yanagisawa}
\affiliation{Institute for Cosmic Ray Research, The University of Tokyo, 5-1-5 Kashiwanoha, Kashiwa, Chiba 277-8582, Japan}
\affiliation{Department of Physics, Graduate School of Science, The University of Tokyo, 7-3-1 Hongo, Bunkyo, Tokyo 113-0033, Japan}
\email[show]{yana@icrr.u-tokyo.ac.jp}

\author[0000-0002-1049-6658]{Masami Ouchi}
\affiliation{National Astronomical Observatory of Japan, 2-21-1 Osawa, Mitaka, Tokyo 181-8588, Japan}
\affiliation{Institute for Cosmic Ray Research, The University of Tokyo, 5-1-5 Kashiwanoha, Kashiwa, Chiba 277-8582, Japan}
\affiliation{Astronomical Science Program, Graduate Institute for Advanced Studies, SOKENDAI, 2-21-1 Osawa, Mitaka, Tokyo 181-8588, Japan}
\affiliation{Kavli Institute for the Physics and Mathematics of the Universe (WPI), The University of Tokyo, 5-1-5 Kashiwanoha, Kashiwa, Chiba 277-8583, Japan}
\email[]{ouchims@icrr.u-tokyo.ac.jp}

\author[orcid=0009-0004-4332-9225]{Tomokazu Kiyota}
\affiliation{National Astronomical Observatory of Japan, 2-21-1 Osawa, Mitaka, Tokyo 181-8588, Japan}
\affiliation{Astronomical Science Program, Graduate Institute for Advanced Studies, SOKENDAI, 2-21-1 Osawa, Mitaka, Tokyo 181-8588, Japan}
\email[]{tomokazu.kiyota@grad.nao.ac.jp}

\author[0009-0004-0381-7216]{Yuta Kageura}
\affiliation{Institute for Cosmic Ray Research, The University of Tokyo, 5-1-5 Kashiwanoha, Kashiwa, Chiba 277-8582, Japan}
\affiliation{Department of Physics, Graduate School of Science, The University of Tokyo, 7-3-1 Hongo, Bunkyo, Tokyo 113-0033, Japan}
\email[]{kageura@icrr.u-tokyo.ac.jp}

\author[0000-0002-4225-4477]{Makoto Ando}
\affiliation{Institute for Cosmic Ray Research, The University of Tokyo, 5-1-5 Kashiwanoha, Kashiwa, Chiba 277-8582, Japan}
\email[]{mando@icrr.u-tokyo.ac.jp}

\author[0000-0002-6047-430X]{Yuichi Harikane}
\affiliation{Institute for Cosmic Ray Research, The University of Tokyo, 5-1-5 Kashiwanoha, Kashiwa, Chiba 277-8582, Japan}
\email[]{hari@icrr.u-tokyo.ac.jp}

\author[0009-0000-1999-5472]{Minami Nakane}
\affiliation{Institute for Cosmic Ray Research, The University of Tokyo,
5-1-5 Kashiwanoha, Kashiwa, Chiba 277-8582, Japan}
\affiliation{Department of Physics, Graduate School of Science, The
University of Tokyo, 7-3-1 Hongo, Bunkyo, Tokyo 113-0033, Japan}
\email[]{nakanem@icrr.u-tokyo.ac.jp}  

\author[0000-0001-9011-7605]{Yoshiaki Ono}
\affiliation{Institute for Cosmic Ray Research, The University of Tokyo, 5-1-5 Kashiwanoha, Kashiwa, Chiba 277-8582, Japan}
\email[]{ono@icrr.u-tokyo.ac.jp}

\author[orcid=0009-0005-2897-002X]{Yui Takeda}
\affiliation{National Astronomical Observatory of Japan, 2-21-1 Osawa, Mitaka, Tokyo 181-8588, Japan}
\email[]{yui.takeda@grad.nao.ac.jp}
\affiliation{Astronomical Science Program, Graduate Institute for Advanced Studies, SOKENDAI, 2-21-1 Osawa, Mitaka, Tokyo 181-8588, Japan}

\begin{abstract}
We investigate the demography of little red dots (LRDs) using 37 objects at $z\sim3$--$7$ with JWST/NIRSpec PRISM and grating spectra compiled from various JWST programs. We focus on spectroscopic quantities of the broad H$\alpha$ luminosity $L_\mathrm{H\alpha,broad}$ (and the broad H$\beta$ luminosity $L_\mathrm{H\beta,broad}$ where available) and the bolometric luminosity $L_\mathrm{bol}$ represented by modified blackbody emission, avoiding quantities contaminated by host-galaxy emission (e.g., total H$\alpha$ luminosity). We identifiy a tight scaling relation between $L_\mathrm{H\alpha,broad}$ and $L_\mathrm{bol}$, supporting the interpretation that these emissions are primarily powered by the central engine. Interestingly, the $L_\mathrm{H\alpha,broad}$–$L_\mathrm{bol}$ scaling relation of LRDs is enhanced by a factor of $\sim40$ in $L_\mathrm{H\alpha,broad}$ relative to that of low-$z$ Type 1 AGN. A similar trend is found in the $L_\mathrm{H\beta,broad}$–$L_\mathrm{bol}$ relation, although the enhancement in $L_\mathrm{H\beta,broad}$ is smaller, only by a factor of $\sim10$. 
We explore the physical origin of these enhancements and find that \textsc{Cloudy} photoionization modeling within the classic locally optimally-emitting cloud (LOC) framework can explain them through an increase in the covering factor from $\sim20$\% (Type 1 AGN) to $\sim100$\% (LRDs), together with an increase in the hydrogen column density from $N_\mathrm{H}\sim10^{23}\,\mathrm{cm}^{-2}$ to $\gtrsim10^{24}\,\mathrm{cm}^{-2}$, with a preferred gas density of $\sim10^{10}\,\mathrm{cm}^{-3}$, successfully reproducing the modified blackbody emission.
Such a nearly unity covering factor without requiring a gas density increase may result from a significant increase in the BLR filling factor or size, corresponding to a ``stuffed BLR" or ``giant BLR," respectively.

\end{abstract}


\section{Introduction}

Recent James Webb Space Telescope (JWST) spectroscopy has revealed a population of compact, broad-line AGN candidates at $z\gtrsim 3$ whose rest-frame UV--optical spectra show a characteristic ``V-shaped'' SED: a blue UV continuum combined with a red optical continuum and often a prominent Balmer break (e.g., \citealp{Kocevski+2023,Harikane+2023,Matthee+2024,Greene+2024,Maiolino+2024,Tanaka+2025,Inayoshi_Ho_2025}). These sources, commonly referred to as little red dots (LRDs), typically show unusually weak X-ray, mid-infrared, and radio emission compared to expectations from classical unobscured AGN, complicating their identification and physical interpretation \citep{Yue+2024,Ananna+2024,Akins+2025,Gloudemans+2025,Maiolino+2025}. The combination of broad Balmer emission and anomalous multi-wavelength properties has motivated scenarios in which the central engine is embedded in optically thick gas (an envelope/cocoon) that can both reprocess the intrinsic emission into a quasi-thermal optical continuum and alter the observed line/continuum ratios \citep{Inayoshi_Maiolino_2025,Inayoshi+2025,Kido+2025}.

Establishing robust empirical relations among observables is essential for testing whether LRDs are powered primarily by accreting black holes and, if so, how their line-emitting regions differ from the broad-line regions (BLRs) of type~1 AGN. In classical AGN, broad Balmer-line luminosities correlate tightly with the optical continuum and with bolometric luminosity, and these relations are widely used as practical calibrations linking recombination-line emission to the ionizing continuum and BLR physics (e.g., \citealp{Greene_Ho_2005,stern_type_2012,Shen_Liu_2012}). If LRDs share the same underlying central engine mechanism, one might expect them to follow similar scaling relations. Conversely, any systematic deviation from the AGN relations would provide direct, quantitative evidence that LRDs have distinct BLR conditions and/or continuum reprocessing.

The total (i.e., narrow$+$broad) Balmer-line luminosities can be significantly affected by host-galaxy emission and narrow-line region contamination, and the limited spectral resolution of JWST/NIRSpec PRISM alone can make decomposition of narrow and broad components uncertain. To isolate the BLR contribution and minimize systematics, it is crucial to measure the luminosity of the broad component using spectra with sufficient resolution. 

This paper is the first in the Archival and Theoretical study of LRDs with AGN comparison across Surveys (ATLAS) series, in which we aim to provide physical interpretations of the LRD population through systematic comparisons with AGNs. Other papers in the series will extend these comparisons to other observational properties, such as Balmer absorption lines  \citep{Yanagisawa+2026_balmerabs}, dust \citep{ Kiyota+2026_LRDdust}, and Ly$\alpha$ emission \citep{Kageura+2026_LRDlya}.
In this paper, we use publicly available JWST/NIRSpec observations to construct a large spectroscopic sample of LRDs with broad H$\alpha$ emission over $z\sim3$--7, combining PRISM spectra with medium- and high-resolution grating data to robustly separate narrow and broad line emission. We then ask two closely related questions: (1) does the long-established AGN $L_{\mathrm{H\alpha,broad}}$--$L_{\mathrm{bol}}$ scaling relation also hold for LRDs, and (2) if so, is it consistent in normalization with type~1 AGN? 
In Section~\ref{sec:data}, we describe the sample construction and data sets. In Section~\ref{sec:analysis}, we detail our spectral fitting and continuum characterization procedures. Section~\ref{sec:results} presents the resulting scaling relations and comparison with previous studies. Section~\ref{sec:discussions} discusses their implications. We summarize our main conclusions in Section~\ref{sec:summary}.
Throughout this paper, we assume cosmology parameters based on the TT, TE, EE + lowE + lensing + BAO result from \cite{Planck+2020} with $H_0=67.66 \, \mathrm{km\,s^{-1}\,Mpc^{-1}}$, $\Omega_\mathrm{m}=0.30966$, and $\Omega_\mathrm{b}=0.04897$. 

\section{Sample and Data}\label{sec:data}
\subsection{DJA Sample}
We use the publicly available JWST/NIRSpec spectra from the Dawn JWST Archive (DJA; \citealt{Heintz+2024, deGraaff+2025a, Valentino+2025}) version~4.4\footnote[1]{\url{https://zenodo.org/records/15472354}}. We start from sources with redshift quality grade 3, and require that a PRISM spectrum is available. To enable consistent emission-line measurements across the sample, we further impose that H$\alpha$ and [O\,\textsc{iii}]$\lambda5007$ are have peak signal-to-noise ratio (SNR) of $\mathrm{SNR}>5$ in at least one of the NIRSpec medium ($R\sim1000$) or high ($R\sim2700$) resolution gratings provided in DJA for each target. The [O\,\textsc{iii}] $\lambda5007$ emission line is used to verify that any outflow-related component has negligible influence on the luminosity measurements of the broad Balmer lines originating from the BLR of LRDs. 
We then search for broad H$\alpha$ emission among the selected targets. 
We perform a double-Gaussian fit consisting of a narrow and a broad component to H$\alpha$. We require the broad component has $\mathrm{FWHM}=700$--$6000\,\mathrm{km\,s^{-1}}$ and $\mathrm{SNR}_{\rm broad}>5.0$. The upper limit on the FWHM is imposed because nearly all targets with $\mathrm{FWHM} > 6000\,\mathrm{km\,s^{-1}}$ do not exhibit a genuine broad H$\alpha$ component; instead, their continua are effectively reproduced by extremely broad Gaussian profiles. After visually inspecting the fitting results, we identify 59 objects that show broad H$\alpha$ emission. 

We further assess the compactness of these objects using NIRCam F444W imaging from the DJA mosaics (version 7\footnote[2]{\url{https://dawn-cph.github.io/dja/imaging/summary/}}; \citealt{Valentino+2023}). Five objects lie outside the NIRCam F444W coverage, 
which are therefore excluded from the subsequent analysis. Aperture photometry is performed using circular apertures with diameters of 0.4$^{\prime\prime}$ and 0.2$^{\prime\prime}$. We adopt the following compactness criterion:

\begin{equation}
    f_\mathrm{F444W}(0.4^{\prime\prime}) / f_\mathrm{F444W}(0.2^{\prime\prime}) < 1.7,
\end{equation}
as employed in e.g., \citet{Labbe+2025} and \citet{Greene+2024}. Here, $f_\mathrm{F444W}$ denotes the F444W flux measured within the specified aperture diameter. We find that only two objects do not satisfy this compactness requirement. We thus exclude these from the sample, resulting in a final set of 52 objects.

\subsection{NIRSpec/IFU Sample}
In addition to the DJA sample, we include an NIRSpec/IFU sample drawn from the programs GO 5665 (PI: Matthee) and GO 5015 (BlackTHUNDER; PI: Übler and Maiolino), 
comprising 8 LRDs and 2 type 1 AGNs. Among these 10, two targets, 159717 and GS-13971, are also selected in DJA sample. Because IFU data have higher SNR, we use IFU data and exclude DJA data for these two objects.

The JWST/NIRSpec IFU data were reduced from the Level-1b uncalibrated products retrieved from MAST. We processed the data with the JWST Calibration Pipeline (v2.0.1) using the CRDS context {\tt\string jwst\_1535.pmap}. For each exposure, we ran the standard Detector1 and Spec2
stages to produce count-rate images and calibrated two-dimensional products. 
We then constructed median-stacked data cubes and corresponding noise cubes from the reduced individual exposures. The resulting spaxel scale was $0.^{\prime\prime}1$. The noise cubes were estimated from the scatter among the contributing frames, scaled by the square root of the number of exposures. The median stacking procedure is described in \cite{Kiyota+2025} in detail.

One-dimensional spectra were extracted from the median-stacked IFU cubes. For each target, the extraction aperture was centered on the source position identified from the H$\alpha$ narrowband map, starting from the catalog coordinates and allowing a local centroid refinement. The source spectrum was then measured within a circular aperture of radius 0.20 arcsec. A local background spectrum was estimated from an annulus with inner and outer radii of 0.45 and 0.95 arcsec, respectively, and subtracted from the source aperture spectrum. The same aperture geometry was also used to extract matched spectra from the PRISM cubes.


\subsection{Sample Summary}
\begin{figure}
    \centering
    \includegraphics[width=1\linewidth]{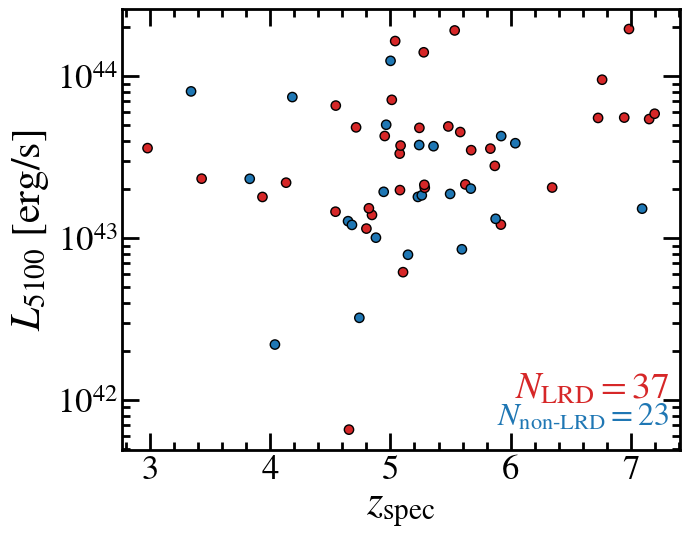}
    \caption{Rest-frame 5100\,\AA\ monochromatic luminosity as a function of spectroscopic redshift for our LRD sample. The red and blue points show the objects classified as LRDs and non-LRDs, respectively.}
    \label{fig:sample}
\end{figure}

Figure~\ref{fig:sample} summarizes the basic properties of our working sample by showing the rest-frame monochromatic luminosity at 5100\,\AA\ ($L_{5100}=\lambda L_\lambda (5100\mathrm{\AA})$) as a function of spectroscopic redshift. We classify each target as an LRD or a non-LRD based on a photometric classification of

\begin{equation}
\begin{aligned}
\mathrm{F277W}-\mathrm{F444W} &> 1.0 \\
\wedge\ \mathrm{F150W}-\mathrm{F200W} &< 1.0 \quad (z<4), \\
\mathrm{F277W}-\mathrm{F444W} &> 1.0 \\
\wedge\ \mathrm{F150W}-\mathrm{F200W} &< 1.0 \quad (4<z<6), \\
\mathrm{F277W}-\mathrm{F444W} &> 1.0 \\
\wedge\ \mathrm{F150W}-\mathrm{F200W} &< 1.0 \quad (z>6).
\end{aligned}
\end{equation}
The criteria for $z>6$ objects are slightly tighter ones than those suggested by \cite{Rinaldi+2026}, and the lower-redshift criteria are similarly defined to select LRDs at $z < 6$. 
For the photometric selection, we create pseudo photometry from the PRISM spectra. After the pseudo-photometric selection, we visually inspect the PRISM spectra to ensure that the red optical color is not mimicked by the strong emission lines. 
Our sample eventually comprises of 37 LRDs and 23 non-LRDs. Both the LRD and non-LRD subsamples span a comparable redshift range of $z_{\rm spec}=3$--7. 
Although our sample is not statistically complete, our sample selection is free from contamination by strong emission lines  and is based on broad-line characteristics, compactness, and continuum-shape criteria. These criteria effectively capture the principal spectral and morphological properties of LRDs or broad-line type 1 AGNs, yielding a high-purity sample rather than a complete one. Tables \ref{tab:sample_objects_lrd} and \ref{tab:sample_objects_nonlrd} summarize the basic information of the sample. 

UNCOVER-13821 and UNCOVER-4286 are gravitationally lensed by the Abell2744 galaxy cluster. For these objects, graviational lensing magnification factors are taken from the UNCOVER magnification catalog v2.0 \citep{Furtak+2023_lensing, Price+2025, Suess+2024, Weaver+2024}. Magnification factors applied for UNCOVER-13821 and UNCOVER-4286 are $\mu = 1.6210_{-0.0148}^{+0.0053}$ and $1.607_{-0.011}^{+0.005}$, respectively.

\begin{deluxetable*}{l l c c c c c c c}
\tablecaption{LRDs in this study \label{tab:sample_objects_lrd}}
\tablehead{
\colhead{ID} & \colhead{Program} & \colhead{RA} & \colhead{Dec} & \colhead{$z_{\rm spec}$} & \colhead{$L_\mathrm{H\alpha,broad}$} & \colhead{$L_\mathrm{H\beta,broad}$} & \colhead{$L_{5100}$} & \colhead{$L_\mathrm{bol}$} \\
\colhead{} & \colhead{} & \colhead{} & \colhead{} & \colhead{} & \colhead{$(10^{42}\,\mathrm{erg\,s^{-1}})$} & \colhead{$(10^{42}\,\mathrm{erg\,s^{-1}})$} & \colhead{$(10^{43}\,\mathrm{erg\,s^{-1}})$} & \colhead{$(10^{43}\,\mathrm{erg\,s^{-1}})$}
}
\startdata
CEERS-746 & CEERS & 14:19:14.19 & +52:52:06.5 & 5.6231 & $2.22^{+0.46}_{-0.32}$ & $\cdots$ & $2.15^{+0.59}_{-0.59}$ & $5.51^{+1.06}_{-0.57}$ \\
EGS-47962 & GO 4106 & 14:19:34.20 & +52:51:24.8 & 6.7273 & $6.95^{+4.71}_{-2.16}$ & $\cdots$ & $5.52^{+0.64}_{-0.64}$ & $11.80^{+2.78}_{-2.19}$ \\
EGS-51623 & GO 4106 & 14:19:32.84 & +52:51:19.4 & 4.9528 & $5.76^{+2.43}_{-1.56}$ & $\cdots$ & $4.27^{+0.24}_{-0.24}$ & $6.88^{+0.11}_{-0.10}$ \\
EGS-71365 & GO 4106 & 14:19:34.93 & +52:54:24.2 & 4.8004 & $0.72^{+0.48}_{-0.26}$ & $\cdots$ & $1.15^{+0.17}_{-0.17}$ & $2.82^{+0.49}_{-0.16}$ \\
JADES-GN-38147 & JADES & 12:37:04.96 & +62:08:54.3 & 5.8683 & $8.81^{+41.02}_{-6.93}$ & $\cdots$ & $2.79^{+0.76}_{-0.76}$ & $18.93^{+11.06}_{-5.90}$ \\
JADES-GN-39353 & JADES & 12:37:10.55 & +62:09:11.1 & 4.8468 & $1.28^{+1.59}_{-0.70}$ & $\cdots$ & $1.39^{+0.41}_{-0.41}$ & $3.49^{+0.61}_{-0.56}$ \\
JADES-GN-68797 & JADES & 12:36:54.99 & +62:08:46.3 & 5.0397 & $42.42^{+9.33}_{-5.32}$ & $10.95^{+3.04}_{-3.03}$ & $16.48^{+0.66}_{-0.66}$ & $95.86^{+4.41}_{-3.55}$ \\
JADES-GN-73488 & JADES & 12:36:47.38 & +62:10:38.0 & 4.1325 & $3.92^{+0.63}_{-0.50}$ & $\cdots$ & $2.20^{+0.13}_{-0.13}$ & $3.87^{+0.04}_{-0.04}$ \\
JADES-GN-53501 & JADES & 12:37:10.81 & +62:11:36.9 & 3.4299 & $4.43^{+0.18}_{-0.26}$ & $\cdots$ & $2.33^{+0.23}_{-0.23}$ & $4.22^{+0.28}_{-0.22}$ \\
JADES-GN-23367 & JADES & 12:36:48.28 & +62:14:56.2 & 2.9800 & $0.69^{+0.10}_{-0.08}$ & $\cdots$ & $3.60^{+0.17}_{-0.17}$ & $2.58^{+0.11}_{-0.15}$ \\
JADES-GN-954 & JADES & 12:36:36.47 & +62:15:34.7 & 6.7608 & $9.49^{+4.26}_{-2.86}$ & $\cdots$ & $9.51^{+0.87}_{-0.87}$ & $15.39^{+0.63}_{-1.14}$ \\
JADES-GS-13704 & JADES & 03:32:30.37 & $-$27:49:05.1 & 5.9189 & $1.63^{+1.27}_{-0.72}$ & $\cdots$ & $1.21^{+0.14}_{-0.14}$ & $4.13^{+0.71}_{-0.51}$ \\
JADES-GS-13329 & JADES & 03:32:33.37 & $-$27:47:04.0 & 3.9358 & $1.39^{+0.86}_{-0.64}$ & $\cdots$ & $1.80^{+0.34}_{-0.34}$ & $2.97^{+0.17}_{-0.15}$ \\
JADES-GS-38562 & JADES & 03:32:32.61 & $-$27:52:17.9 & 4.8210 & $2.36^{+0.93}_{-0.56}$ & $\cdots$ & $1.53^{+0.20}_{-0.20}$ & $3.90^{+0.28}_{-0.34}$ \\
RUBIES-EGS-925921 & RUBIES & 14:20:18.01 & +52:56:37.8 & 6.9448 & $12.65^{+1.96}_{-1.70}$ & $\cdots$ & $5.55^{+2.13}_{-2.13}$ & $26.29^{+36.54}_{-9.84}$ \\
RUBIES-EGS-926125 & RUBIES & 14:20:32.90 & +52:59:18.8 & 5.2858 & $3.99^{+0.85}_{-0.86}$ & $\cdots$ & $2.05^{+0.49}_{-0.49}$ & $5.86^{+2.87}_{-0.53}$ \\
RUBIES-EGS-24948 & RUBIES & 14:19:55.52 & +52:52:57.2 & 4.5457 & $0.64^{+0.82}_{-0.51}$ & $\cdots$ & $6.58^{+0.68}_{-0.68}$ & $10.88^{+0.72}_{-0.62}$ \\
RUBIES-EGS-29489 & RUBIES & 14:20:05.30 & +52:55:14.8 & 4.5434 & $2.52^{+1.20}_{-0.59}$ & $\cdots$ & $1.46^{+0.42}_{-0.42}$ & $1.96^{+0.63}_{-0.48}$ \\
RUBIES-EGS-42046 & RUBIES & 14:19:10.89 & +52:47:19.8 & 5.2772 & $30.82^{+14.32}_{-12.82}$ & $3.96^{+2.76}_{-2.09}$ & $14.04^{+0.68}_{-0.68}$ & $19.33^{+1.95}_{-2.29}$ \\
RUBIES-EGS-55604 & RUBIES & 14:19:55.93 & +52:57:21.6 & 6.9843 & $55.99^{+14.37}_{-12.94}$ & $8.39^{+3.61}_{-3.48}$ & $19.55^{+1.86}_{-1.86}$ & $36.11^{+2.70}_{-1.41}$ \\
RUBIES-EGS-61496 & RUBIES & 14:19:53.39 & +52:57:43.9 & 5.0795 & $2.27^{+0.98}_{-0.77}$ & $\cdots$ & $1.98^{+0.69}_{-0.69}$ & $3.24^{+0.56}_{-0.25}$ \\
RUBIES-EGS-952625 & RUBIES & 14:19:54.13 & +52:55:31.0 & 5.1051 & $0.44^{+0.84}_{-0.33}$ & $\cdots$ & $0.62^{+0.76}_{-0.76}$ & $3.20^{+0.96}_{-0.44}$ \\
RUBIES-EGS-42803 & RUBIES & 14:19:43.09 & +52:53:16.5 & 7.1522 & $7.08^{+7.55}_{-3.69}$ & $\cdots$ & $5.43^{+1.61}_{-1.61}$ & $11.32^{+2.31}_{-1.57}$ \\
RUBIES-UDS-19484 & RUBIES & 02:16:55.78 & $-$05:16:50.4 & 4.6555 & $0.73^{+0.99}_{-0.43}$ & $\cdots$ & $0.07^{+0.48}_{-0.48}$ & $3.32^{+0.32}_{-0.35}$ \\
RUBIES-UDS-182791 & RUBIES & 02:16:51.32 & $-$05:05:13.4 & 4.7145 & $10.09^{+6.97}_{-4.67}$ & $\cdots$ & $4.83^{+0.42}_{-0.42}$ & $6.55^{+1.20}_{-0.99}$ \\
RUBIES-UDS-970351 & RUBIES & 02:17:02.86 & $-$05:06:18.7 & 5.2822 & $2.36^{+2.36}_{-1.18}$ & $\cdots$ & $2.13^{+0.59}_{-0.59}$ & $3.38^{+0.84}_{-0.30}$ \\
RUBIES-UDS-172350 & RUBIES & 02:17:28.55 & $-$05:06:14.2 & 5.5810 & $7.55^{+4.32}_{-3.29}$ & $\cdots$ & $4.52^{+0.65}_{-0.65}$ & $7.59^{+0.58}_{-0.63}$ \\
RUBIES-UDS-47509 & RUBIES & 02:17:03.50 & $-$05:13:57.3 & 5.6718 & $4.48^{+0.51}_{-0.76}$ & $\cdots$ & $3.49^{+0.82}_{-0.82}$ & $8.52^{+1.40}_{-0.75}$ \\
RUBIES-UDS-981721 & RUBIES & 02:17:29.37 & $-$05:06:57.3 & 7.1977 & $12.16^{+3.58}_{-2.60}$ & $\cdots$ & $5.86^{+1.94}_{-1.94}$ & $14.66^{+3.03}_{-4.04}$ \\
159717 & BlackThunder & 03:32:23.40 & $-$27:54:04.7 & 5.0775 & $7.19^{+0.54}_{-0.47}$ & $\cdots$ & $3.32^{+0.28}_{-0.28}$ & $14.31^{+0.91}_{-1.04}$ \\
UNCOVER-13821
& BlackThunder & 00:14:28.94 & $-$30:24:00.0 & 6.3454 & $5.87^{+0.58}_{-0.53}$ & $\cdots$ & $2.05^{+1.54}_{-1.54}$ & $7.56^{+0.92}_{-1.44}$ \\
UNCOVER-4286
& BlackThunder & 00:14:28.61 & $-$30:25:23.9 & 5.8315 & $3.82^{+0.47}_{-0.47}$ & $\cdots$ & $3.57^{+0.85}_{-0.85}$ & $3.80^{+0.09}_{-0.18}$ \\
J1148-18404 & BlackThunder & 11:48:13.92 & +52:51:46.1 & 5.0115 & $7.10^{+0.49}_{-0.50}$ & $\cdots$ & $7.14^{+0.86}_{-0.86}$ & $15.27^{+0.17}_{-0.12}$ \\
GN-12839 & GO 5664 & 12:37:22.75 & +62:15:47.2 & 5.2411 & $24.02^{+0.21}_{-0.24}$ & $\cdots$ & $4.80^{+0.30}_{-0.30}$ & $27.06^{+0.52}_{-0.54}$ \\
GN-15498 & GO 5664 & 12:37:08.52 & +62:16:50.9 & 5.0846 & $4.64^{+0.75}_{-0.65}$ & $\cdots$ & $3.73^{+0.30}_{-0.30}$ & $7.27^{+0.07}_{-0.04}$ \\
GN-9771 & GO 5664 & 12:37:07.44 & +62:14:50.3 & 5.5346 & $52.07^{+0.97}_{-1.01}$ & $5.98^{+1.28}_{-0.95}$ & $19.16^{+0.43}_{-0.43}$ & $21.56^{+0.11}_{-0.08}$ \\
GS-13971 & GO 5664 & 03:32:33.26 & $-$27:47:25.1 & 5.4816 & $6.81^{+0.09}_{-0.10}$ & $\cdots$ & $4.89^{+0.31}_{-0.31}$ & $9.57^{+0.69}_{-0.27}$ \\
\enddata
\tablecomments{Program references: CEERS \citep{Finkelstein+2022}; GO-4106 (PI: Nelson); JADES \citep{Eisenstein+2026}; RUBIES \citep{deGraaff+2025a}; BlackTHUNDER \citep{Ubler+2025}; \citep{Torralba+2026, Matthee+2026}.}
\end{deluxetable*}
\begin{deluxetable*}{l l c c c c c}
\tablecaption{non-LRDs in this study \label{tab:sample_objects_nonlrd}}
\tablehead{
\colhead{ID} & \colhead{Program} & \colhead{RA} & \colhead{Dec} & \colhead{$z_{\rm spec}$} & \colhead{$L_\mathrm{H\alpha,broad}$} & \colhead{$L_{5100}$} \\
\colhead{} & \colhead{} & \colhead{} & \colhead{} & \colhead{} & \colhead{$(10^{42}\,\mathrm{erg\,s^{-1}})$} & \colhead{$(10^{43}\,\mathrm{erg\,s^{-1}})$}
}
\startdata
JADES-GN-77652 & JADES & 12:37:10.37 & +62:11:56.4 & 5.2286 & $1.72^{+1.09}_{-0.56}$ & $1.80^{+0.59}_{-0.59}$ \\
JADES-GN-61888 & JADES & 12:36:40.32 & +62:13:01.2 & 5.8752 & $1.67^{+2.29}_{-1.27}$ & $1.32^{+0.33}_{-0.33}$ \\
JADES-GN-1093 & JADES & 12:36:43.14 & +62:13:28.7 & 5.5944 & $1.21^{+2.80}_{-0.87}$ & $0.85^{+0.37}_{-0.37}$ \\
JADES-GN-20621 & JADES & 12:36:29.40 & +62:17:34.3 & 4.6802 & $1.25^{+17.02}_{-1.19}$ & $1.21^{+0.46}_{-0.46}$ \\
JADES-GN-1121 & JADES & 12:36:31.16 & +62:16:52.2 & 3.3420 & $1.25^{+11.36}_{-1.13}$ & $8.05^{+0.21}_{-0.21}$ \\
JADES-GS-9515 & JADES & 03:32:31.88 & $-$27:48:06.7 & 4.6477 & $1.32^{+0.11}_{-0.09}$ & $1.27^{+0.08}_{-0.08}$ \\
JADES-GS-13520 & JADES & 03:32:31.58 & $-$27:48:35.6 & 4.9441 & $1.12^{+1.09}_{-0.62}$ & $1.93^{+0.26}_{-0.26}$ \\
JADES-GS-10013268 & JADES & 03:32:48.44 & $-$27:49:15.2 & 4.0395 & $0.31^{+0.47}_{-0.16}$ & $0.22^{+0.18}_{-0.18}$ \\
JADES-GS-179198 & JADES & 03:32:21.35 & $-$27:51:38.5 & 3.8299 & $0.77^{+0.29}_{-0.24}$ & $2.32^{+0.15}_{-0.15}$ \\
JADES-GS-172975 & JADES & 03:32:21.06 & $-$27:52:16.5 & 4.7417 & $0.88^{+0.14}_{-0.11}$ & $0.32^{+0.26}_{-0.26}$ \\
JADES-GS-73690 & JADES & 03:32:14.53 & $-$27:50:54.2 & 5.4967 & $1.11^{+1.13}_{-0.62}$ & $1.87^{+0.38}_{-0.38}$ \\
JADES-GS-30148179 & JADES & 03:32:34.10 & $-$27:46:47.5 & 5.9214 & $2.07^{+2.56}_{-0.48}$ & $4.27^{+0.47}_{-0.47}$ \\
JADES-GS-165174 & JADES & 03:32:26.42 & $-$27:53:08.7 & 4.1843 & $2.01^{+0.20}_{-0.28}$ & $7.43^{+0.13}_{-0.13}$ \\
RUBIES-EGS-13872 & RUBIES & 14:20:31.90 & +52:58:14.5 & 5.2611 & $1.20^{+1.11}_{-0.56}$ & $1.84^{+0.50}_{-0.50}$ \\
RUBIES-EGS-6411 & RUBIES & 14:20:26.20 & +52:56:23.2 & 4.8796 & $0.55^{+2.04}_{-0.41}$ & $1.01^{+0.39}_{-0.39}$ \\
RUBIES-EGS-46985 & RUBIES & 14:19:13.36 & +52:48:34.2 & 4.9649 & $2.96^{+2.45}_{-1.71}$ & $5.01^{+6.99}_{-6.99}$ \\
RUBIES-EGS-50052 & RUBIES & 14:19:17.63 & +52:49:49.0 & 5.2394 & $2.47^{+3.29}_{-0.79}$ & $3.76^{+0.41}_{-0.41}$ \\
RUBIES-EGS-17146 & RUBIES & 14:19:47.88 & +52:50:43.5 & 5.0012 & $1.73^{+3.67}_{-1.27}$ & $12.43^{+1.11}_{-1.11}$ \\
RUBIES-UDS-174752 & RUBIES & 02:16:49.39 & $-$05:06:01.8 & 6.0389 & $3.51^{+3.49}_{-1.87}$ & $3.86^{+0.86}_{-0.86}$ \\
RUBIES-UDS-920396 & RUBIES & 02:16:59.48 & $-$05:09:07.2 & 7.0935 & $7.15^{+1.82}_{-1.15}$ & $1.52^{+2.77}_{-2.77}$ \\
RUBIES-UDS-19521 & RUBIES & 02:17:32.08 & $-$05:17:15.8 & 5.6687 & $1.53^{+4.25}_{-1.19}$ & $2.02^{+0.83}_{-0.83}$ \\
GN-16813 & GO 5664 & 12:36:43.03 & +62:17:33.0 & 5.3585 & $4.87^{+0.18}_{-0.33}$ & $3.70^{+0.35}_{-0.35}$ \\
GN-14409 & BlackThunder & 12:36:17.30 & +62:16:24.2 & 5.1461 & $4.44^{+0.07}_{-0.08}$ & $0.79^{+0.42}_{-0.42}$ \\
\enddata
\tablecomments{Program references: JADES \citep{Eisenstein+2026}; RUBIES \citep{deGraaff+2025a}; BlackTHUNDER \citep{Ubler+2025}; GO~5664 \citep{Torralba+2026, Matthee+2026}.}
\end{deluxetable*}

\begin{figure*}
    \centering
    \includegraphics[width=1\linewidth]{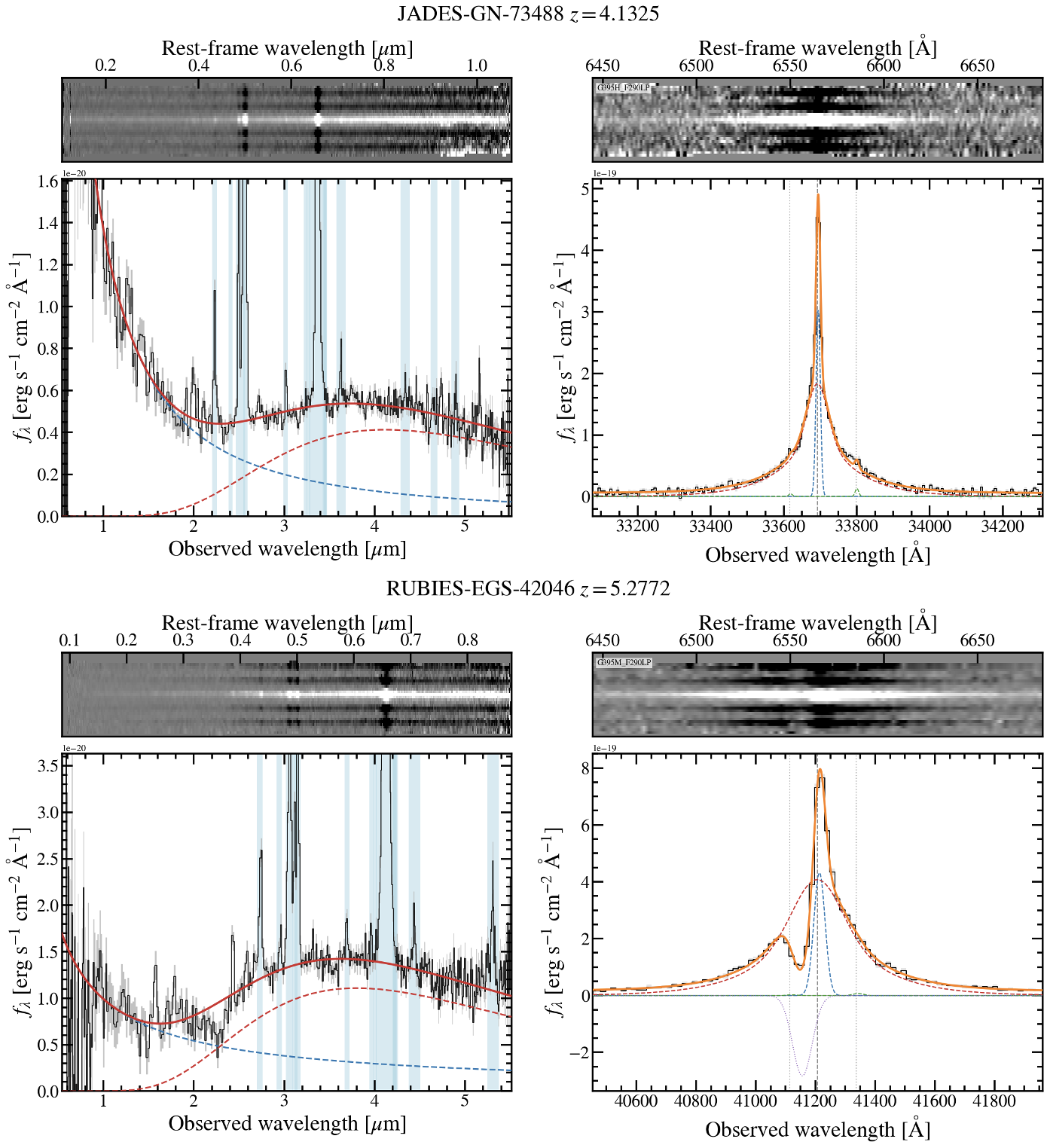}
    \caption{Representative examples of our spectral analysis. For each object (top: JADES-GN-73488; bottom: RUBIES-EGS-42046), the left panels show the NIRSpec/PRISM spectrum (black) together with the best-fit modified blackbody and power law UV continuum model (red solid curve). The red and blue dashed curves indicate the modified blackbody and power law continuum components, respectively. The blue shaded regions indicate masked intervals around emission lines. The right panels show the fit to the H$\alpha$+[N\,\textsc{ii}] complex using the grating spectrum (black), decomposed into a broad component (red), narrow component (blue), [N\,\textsc{ii}] lines (green), absorption line (purple), and the total model (orange) as described in Section~\ref{sec:line_profile_fitting}. The vertical dotted lines show the rest-frame wavelengths of [N \textsc{ii}]$\lambda\lambda6548,6583$ and H$\alpha$.}
    \label{fig:example_spectra}
\end{figure*}

\section{Analysis}\label{sec:analysis}
\subsection{Line Profile Fitting}\label{sec:line_profile_fitting}
Using grating spectra, we fit the H$\alpha$+[N\,\textsc{ii}] complex in the DJA and NIRSpec/IFU samples. If the H$\alpha$ line is covered by both medium- and high-resolution grating, we use high-resolution one. We fit the H$\alpha$+[N\,\textsc{ii}] complex with the following components:

\begin{itemize}
    \item Narrow emission component: a single Gaussian with $\mathrm{FWHM}<\mathrm{500}\,\mathrm{km\,s^{-1}}$.
    \item Broad emission component: an exponential profile with an FWHM broader than the narrow Gaussian. The adoption of an exponential profile is motivated by several recent studies, which demonstrate that most high–signal-to-noise LRDs are better described by an exponential rather than a Gaussian profile \citep[e.g.,][]{Rusakov+2026, Scholtz+2026}.
    \item [$\bullet$] [N\,\textsc{ii}] $\lambda\lambda6548,6583$: each line is modeled as a single Gaussian, with the line ratio fixed to the theoretical value of 2.94 \citep{Galavis+1997}.
    \item Balmer absorption (optional): we perform two fits for each object, one without absorption and one including an additional absorption component modeled as a single Gaussian.
    \item Continuum: a linear continuum is assumed.
\end{itemize}

We adopt a nominal rest-frame wavelength interval of $6564\pm 120\,\mathrm{\AA}$ to ensure that the full extent of the broad emission line is encompassed by the fit. In cases where evident artifacts are present within this interval, the affected regions are excluded from the fitting range. We convolve these components with line spread functions obtained by \cite{Isobe+2023}. When available, we also fit the H$\beta$ emission line in the same manner, except that the [N\,\textsc{ii}] doublet is not included. We infer the posterior distributions of the line-profile parameters using a Markov Chain Monte Carlo (MCMC) approach implemented with the {\tt emcee} ensemble sampler \citep{Foreman-Mackey+13}. Unless otherwise stated, we adopt the 50th percentile of the marginalized posterior as the best-fit value, and the 16th and 84th percentiles as the corresponding uncertainties. To decide whether Balmer absorption is required, we fit each spectrum both with and without the absorption component and compute the Bayesian Information Criterion (BIC) for each model. We then define $\Delta\mathrm{BIC}\equiv \mathrm{BIC}_{\rm no\,abs}-\mathrm{BIC}_{\rm abs}$ and adopt the absorption-included solution when $\Delta\mathrm{BIC}>15$; otherwise we retain the no-absorption fit. Representative examples of the fits are shown in Figure \ref{fig:example_spectra}.

\subsection{Continuum Characterization}\label{sec:continuum}
We characterize the continuum shape using the JWST/NIRSpec PRISM spectra. As a first-order description, we model the continuum with a modified blackbody as done by \cite{deGraaff+2025c}:

\begin{equation}
    f_\nu \propto B_\nu(T_\mathrm{BB})\left(\nu/\nu_0\right)^{\beta_\mathrm{MBB}},
\end{equation}
where $B_\nu(T_\mathrm{BB})$ is the Planck function with temperature of $T_\mathrm{BB}$, $\nu_0$ is the pivot frequency $\nu_0 = c/(5500 \mathrm{\AA}) = 5.45\times 10^{14} \mathrm{Hz}$, and $\beta_\mathrm{MBB}$ is the power law index of modification. The adoption of a modified blackbody instead of a pure blackbody is motivated by indications of nebular continuum contributing to the optical continua of LRDs \citep[e.g.,][]{Sneppen+2026b}. In such cases, the continuum shape could depart from that of a pure blackbody due to the contribution from the nebular continuum.
However, because a detailed characterization of the physical origin of the continuum shape is beyond the scope of this study, our objective is instead to reproduce the overall spectral profile and to infer the continuum luminosity using this simplified parameterization.

Following suggestions that the rest-frame UV continuum of LRDs is spatially extended and may be dominated by host-galaxy emission rather than an unresolved central source \citep[e.g.,][]{Rinaldi+2025,Cloonan+2026}, we include an additional power-law component, $f_\lambda \propto \lambda^{\beta_\mathrm{UV}}$, where $\beta_\mathrm{UV}$ is the UV spectral slope. We then fit the modified-blackbody and power-law components simultaneously to the full PRISM spectrum. 
We mask wavelength intervals around prominent emission lines (H$\gamma$, He \textsc{ii} $\lambda$4686, H$\beta$, [O \textsc{iii}]$\lambda\lambda$4959,5007, He \textsc{i} $\lambda$5876, [O \textsc{i}]$\lambda\lambda$6300,6364, H$\alpha$, [N \textsc{ii}]$\lambda\lambda6548,6583$, [S \textsc{ii}]$\lambda\lambda$6718,6732, and He \textsc{i} $\lambda$7065) so that the inferred continuum parameters are not biased by line flux. 
We do not perform continuum fitting for the non-LRD sample because their optical continua are not adequately described by the modified blackbody model. Instead, for non-LRDs we only use $L_\mathrm{5100}$.



\begin{figure*}
    \centering
    \includegraphics[width=1\linewidth]{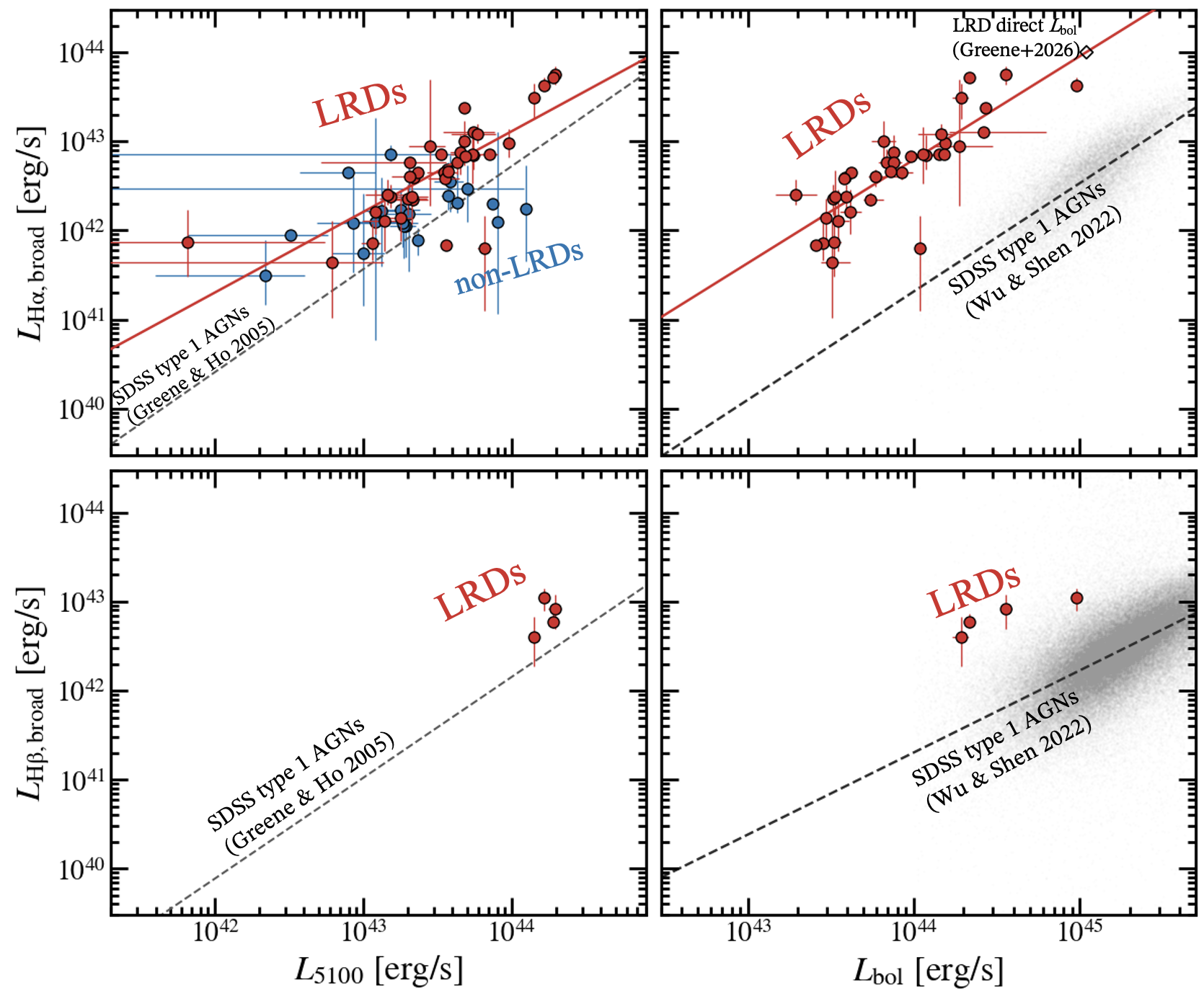}
    \caption{Scaling relations between broad Balmer line luminosities and continuum luminosities. The top and bottom rows present $\lbha$ and $L_\mathrm{H\beta,broad}$, respectively, while the left and right column show $L_{5100}$ and $\lbol$, respectively. The red and blue points show luminosities of LRDs and non-LRDs measured in this work, respectively. The red solid line indicates our best-fit relation to the LRD data. The gray points and black lines show measurements and their best-fit relations for type 1 AGNs $z\lesssim2$ from \citet{Greene_Ho_2005} (left) and \citet{Wu_Shen_2022} (right). The open diamond indicates the direct $\lbol$ measurement of Abell2744-45924 \citep{Greene+2025}.}
    \label{fig:scaling_relation}
\end{figure*}

\begin{figure*}
    \centering
    \includegraphics[width=1\linewidth]{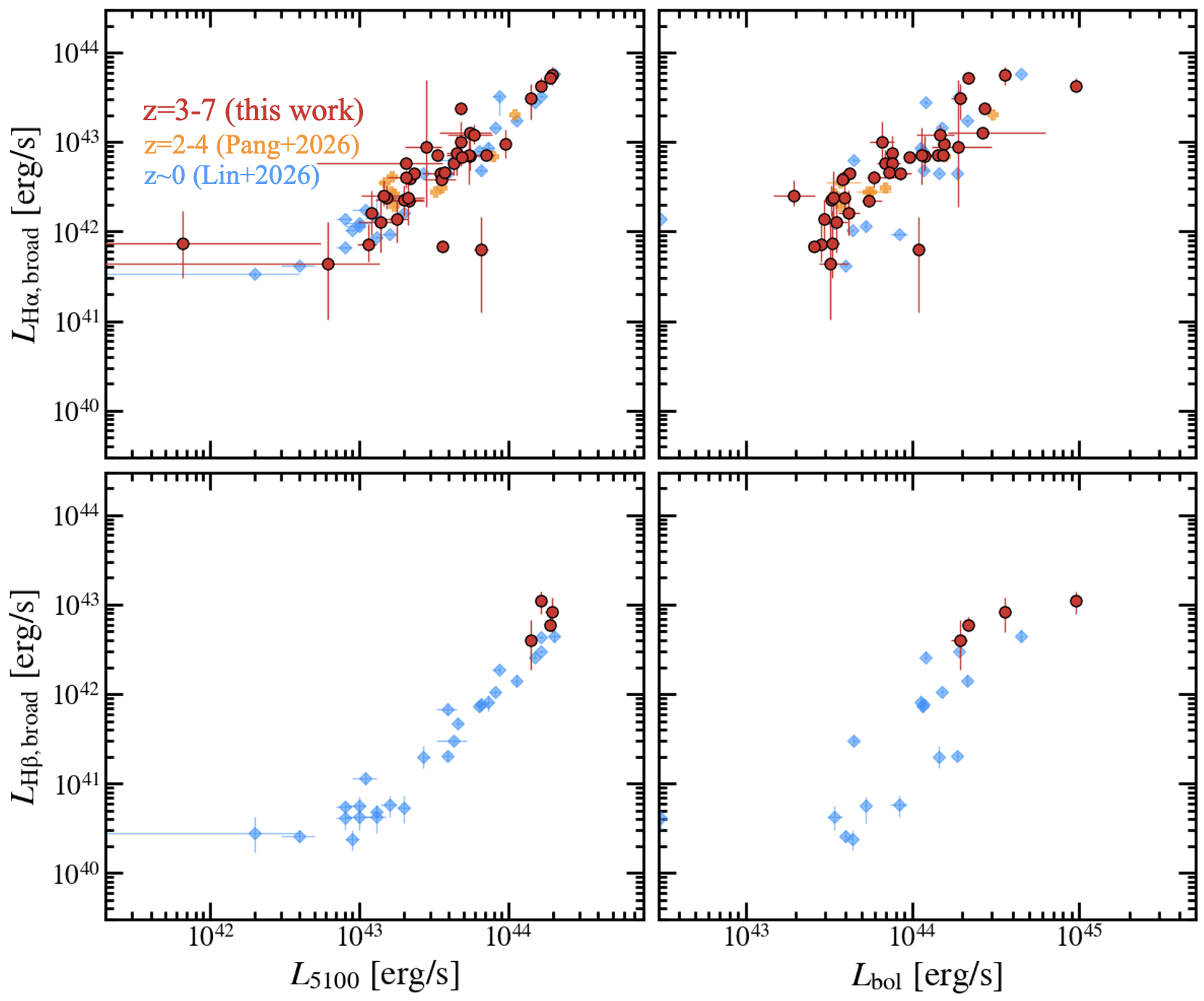}
    \caption{Same as Figure \ref{fig:scaling_relation}, but comparing the LRDs at $z=3$--7 in this work (red circles), those at $z\sim0$ from \cite{XLin+2026} (blue diamonds), and those at $z\sim2$--4 from \cite{Pang+2026} (orange crosses).}
    \label{fig:scaling_relation_lin}
\end{figure*}

\begin{figure*}
    \centering
    \includegraphics[width=1\linewidth]{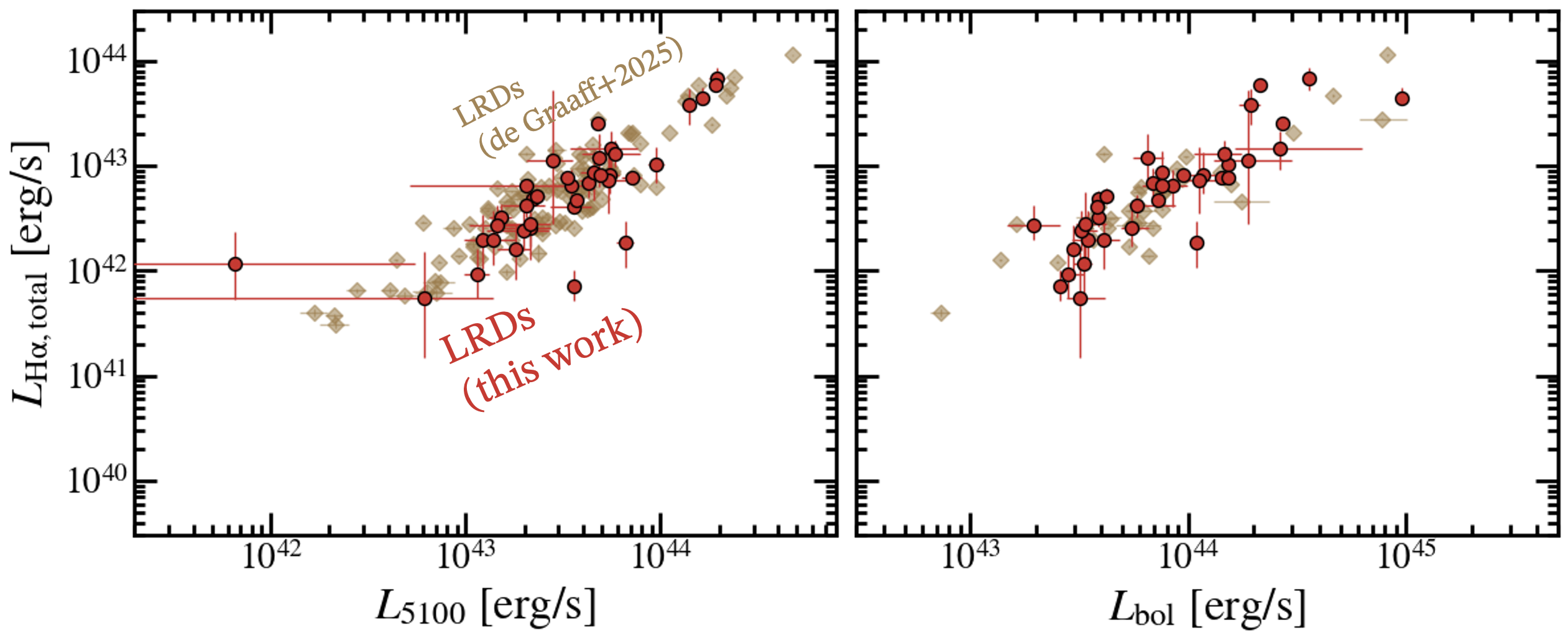}
    \caption{Same as Figure \ref{fig:scaling_relation}, but comparing the LRDs with total H$\alpha$ measurements in this work (red circles) and those from \cite{deGraaff+2025c} (brown diamonds).}
    \label{fig:scaling_relation_degraaff}
\end{figure*}

\section{Results}\label{sec:results}
The top-left panel of Figure~\ref{fig:scaling_relation} presents the empirical relation between the broad H$\alpha$ luminosity, $\lbha$, and $L_{5100}$. We identify a correlation between the two quantities, and a power-law fit yields
\begin{equation}
\begin{split}
\log\left(\frac{\lbha}{\mathrm{erg\,s^{-1}}}\right)
&= (0.91^{+0.13}_{-0.13})\,\times\,\log\left(\frac{L_{5100}}{\mathrm{erg\,s^{-1}}}\right) - 3.2^{+5.7}_{-5.5},\\
\sigma_\mathrm{int} &= 0.35^{+0.05}_{-0.04}.
\end{split}
\end{equation}
The correlation between $\lbha$ and $\lbol$ indicates that both broad line and modified-blackbody continuum are likely to be powered by the same energy source. Relative to the SDSS type~1 AGN scaling relation of \citet{Greene_Ho_2005}, our best-fit relation implies that LRDs are systematically brighter in broad H$\alpha$ by $\sim$0.5--1\,dex at fixed $L_{5100}$. 
We also find that the non-LRD broad line AGNs lie between LRDs and low-$z$ type 1 AGNs \citep{Greene_Ho_2005} with larger scatters, suggesting that non-LRD broad line AGNs at high-$z$ might be a transitioning phase from LRDs to low-$z$ type 1 AGNs.



Because our sample requires a detected broad H$\alpha$ component, the selection may preferentially include sources with larger $L_{\mathrm{H\alpha,broad}}$, potentially enhancing the observed offset from the low-$z$ type~1 AGN scaling relation. However, the offset persists even at the bright end of our sample, where incompleteness against low-$L_{\mathrm{H\alpha,broad}}$ LRDs is expected to be least severe. In particular, LRDs with $L_{\mathrm{H\alpha,broad}}\gtrsim10^{43}\,\mathrm{erg\,s^{-1}}$ still lie systematically above the low-$z$ type~1 AGN locus. Therefore, even if our broad-line selection introduces a bias in the full distribution of $L_{\mathrm{H\alpha,broad}}$, the conclusion that the luminous systems exhibit an intrinsic offset is robust.

In the bottom-left panel of Figure \ref{fig:scaling_relation}, we also find that four of the LRDs, which exhibit broad H$\beta$ emission lines, show a similar yet smaller offset from the scaling relation reported by \citet{Greene_Ho_2005}.

In the top-right panel of Figure~\ref{fig:scaling_relation}, we additionally show the relation between $\lbha$ and the bolometric luminosity, $L_{\rm bol}$, under the working assumption that the fitted modified-blackbody luminosity provides a proxy for the intrinsic bolometric output of LRDs \citep{Inayoshi+2025, Umeda+2025}. Under this assumption, our inferred $L_\mathrm{H\alpha,broad}$--$L_{\rm bol}$ relation is consistent with the direct $\lbol$ measurements reported by \citet{Greene+2025} for 
a luminous LRD, Abell2744-45924 \citep{Labbe+2024, Setton+2025}, constrained by broad-band observations from X-ray through far-infrared. At the same time, we find that LRDs remain offset from the SDSS type~1 AGN relation of \citet{Wu_Shen_2022}, with broad H$\alpha$ being brighter by a factor of $\sim40$ at fixed $L_{\rm bol}$. We find a linear relation of

\begin{equation}
\begin{split}
\log\left(\frac{\lbha}{\mathrm{erg\,s^{-1}}}\right)
&= (1.2^{+0.1}_{-0.1})\,\times\,\log\left(\frac{\lbol}{\mathrm{erg\,s^{-1}}}\right) - 8.3^{+6.0}_{-6.0},\\
\sigma_\mathrm{int} &= 0.31^{+0.04}_{-0.03}.
\end{split}
\end{equation}
The broad H$\beta$ also show the offset from the relation of \citet{Wu_Shen_2022}, but the offset is smaller than that of H$\alpha$ (by a factor of $\sim10$; the bottom-right panel of Figure \ref{fig:scaling_relation}). 

In Figure~\ref{fig:scaling_relation_lin}, we compare our measurements with those of $z\sim0$ LRDs from \citet{XLin+2026} and $z=2$--4 LRDs from \cite{Pang+2026}. Their lower redshift measurements fall on the same locus as our higher redshift sample, suggesting that the scaling relation of LRDs may exhibit no redshift evolution and could represent a nearly universal scaling for LRDs across cosmic time. Figure~\ref{fig:scaling_relation_degraaff} further compares our measurements of total H$\alpha$ luminosity, $L_\mathrm{H\alpha,total}$, with those of \citet{deGraaff+2025c}, which are based on the JWST/NIRSpec PRISM spectra. We find that our measurent and those of \citet{deGraaff+2025c} are broadly consistent with each other. 

\section{Discussions}\label{sec:discussions}
\subsection{Locally Optimally-emitting Cloud (LOC) Model}
To interpret the empirical scaling relations of LRDs found in the previous section, we employ photoionization modeling. As a conceptual framework, we introduce the locally optimally-emitting cloud (LOC) model \citep{Baldwin+1995,Korista+1997,Korista_Goad_2004}. The LOC model is a classical approach originally proposed to describe the BLR of AGNs, in which the line-emitting gas spans a broad range of hydrogen number density $n_\mathrm{H}$ and ionization parameter $U$. A key idea is that, for a given observed emission line, the emergent luminosity is dominated by the subset of clouds whose emissivity in that line is near its maximum in the $(n_\mathrm{H},U)$ plane, while contributions from clouds away from the optimum conditions are sub-dominant.
Motivated by this concept, we focus on gas conditions in the vicinity of the optimal $(n_\mathrm{H},U)$ values to connect the observed $L_\mathrm {H\alpha,broad}$ scaling relations to the underlying ionizing continuum and BLR gas properties. Although their LOC models employ fine grids of calculations and take a weighted sum of them around the optimal gas conditions, we focus on 
a single $(n_\mathrm{H},U)$ pair of values (i.e., one-zone model) for simplicity. 

\subsection{Cloudy Setup}
We model the BLR gas with the photoionization code \textsc{Cloudy} version 23 \citep{Gunasekera+2023}. We adopt a 
standard AGN ionizing spectrum normalized to a bolometric luminosity of $L_\mathrm{bol}=10^{42-46}\,\mathrm{erg\,s^{-1}}$. Motivated by a broad line width of Balmer lines, we set a gas turbulent velocity of $300\,\mathrm{km\,s^{-1}}$, which is also used in \textsc{Cloudy} modeling of MoM-BH*1 \citep{Naidu+2025}. Gas phase metallicity is set to 0.1 $Z_\odot$. Guided by the LOC arguments above, we fix the gas density and ionization parameter to $\log(n_\mathrm{H}/\mathrm{cm}^{-3})=10$ and $\log U=-2$, which are standard BLR parameters for the single-zone model \citep[e.g.,][]{Davidson_Netzer_1979, Peterson+2006}. 

With $n_\mathrm{H}$ and $U$ held fixed, changing the ionizing photon luminosity alone does not generally reproduce the enhanced Balmer-line luminosities at given $\lbol$; instead, the Balmer-line output at a given ionizing continuum level primarily scales with the total amount of Balmer line emitting gas. We therefore treat the gas covering factor, $f_{\rm cov}$, and the hydrogen column density, $N_{\rm H}$, as free parameters that control the effective emitting area and depth of the BLR, respectively. We use a plane-parallel geometry with $f_\mathrm{cov}=0.2$, which is a typical BLR gas covering factor of type 1 AGNs \citep[e.g.,][]{Baldwin+1995,Korista_Goad_2004}, and closed geometry as a $f_\mathrm{cov}=1$ case. Here we specify {\tt\string sphere expanding} command in \textsc{Cloudy} for the closed geometry. We change $\log (N_{\rm H}/\mathrm{cm^{-2}})$ from 22 to 25. The \textsc{Cloudy} setup adopted in this work is summarized in Table~\ref{tab:cloudy_setup}.

\begin{deluxetable*}{lc}
\tablecaption{Summary of the \textsc{Cloudy} setup\label{tab:cloudy_setup}}
\tablehead{\colhead{Parameter} & \colhead{Value}}
\startdata
Geometry & plane parallel ($f_{\rm cov}=0.2$) or closed geometry ($f_{\rm cov}=1$) \\
Incident radiation & AGN \\
$\log (L_\mathrm{bol}/\mathrm{erg~s^{-1}})$ & 42--46 \\
Turbulent velocity & $300\,\mathrm{km\,s^{-1}}$ (fixed) \\
Metallicity & 0.1 $Z_\odot$ (fixed) \\
$\log(n_\mathrm{H}/\mathrm{cm}^{-3})$ & $\mathrm{10}$ (fixed) \\
$\log U$ & $\mathrm{-2}$ (fixed) \\
$\log (N_\mathrm{H}/\mathrm{cm^{-2}})$ & 22--26 \\
\enddata
\end{deluxetable*}

\subsection{Cloudy Modeling Result:\\ ``Stuffed BLR" or ``Giant BLR"?}
\begin{figure*}
    \centering
    \includegraphics[width=1\linewidth]{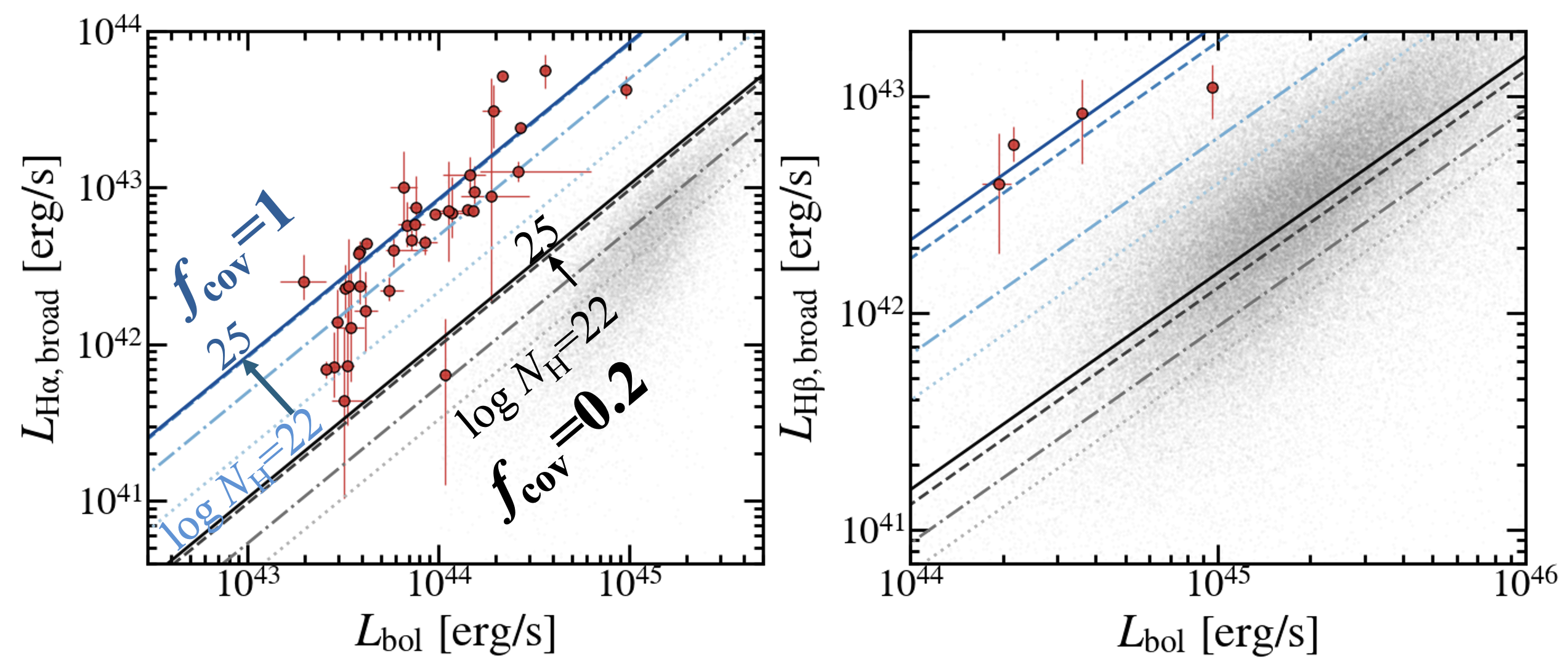}
    \caption{Comparison of the observed broad Balmer-line luminosities and bolometric luminosity with \textsc{Cloudy} photoionization models. Left: $\lbha$ versus $L_{\rm bol}$. Right: $L_\mathrm{H\beta,broad}$ versus $L_{\rm bol}$. The symbols are the same as Figures \ref{fig:scaling_relation}, \ref{fig:scaling_relation_lin}, and \ref{fig:scaling_relation_degraaff}. The gray and black lines show the models with $f_\mathrm{cov}=0.2$, while the blue lines show those with $f_\mathrm{cov}=1$. The dotted, dot-dashed, dashed, and solid lines correspond to $\log (N_\mathrm{H}/\mathrm{cm^{-2}}) = 22, 23, 24,$ and 25, respectively.}
    \label{fig:cloudy}
\end{figure*}

Figure~\ref{fig:cloudy} compares the observed scaling relation to the predictions of our \textsc{Cloudy} models. For a covering factor of $f_{\rm cov}=0.2$, we vary the hydrogen column density from $\log(N_{\rm H}/\mathrm{cm}^{-2})=22$ to 25. Across this range, the predicted values of $L_{\rm H\alpha,broad}$ and $L_{\rm H\beta,broad}$ at a given $\lbol$ increase only modestly, and does not reach the elevated locus occupied by LRDs. Instead, the $f_{\rm cov}=0.2$ sequence is broadly consistent with the low-$z$ type~1 AGN relation. 
In contrast, increasing the covering factor to $f_{\rm cov}=1$ substantially boosts the Balmer-line output. In particular, models with $f_{\rm cov}=1$ and $\log(N_{\rm H}/\mathrm{cm}^{-2})\sim24$--25 reproduce the observed LRD relation. 
Although $f_\mathrm{cov}$ and $N_\mathrm{H}$ are partially degenerate, the observed relation requires both parameters to be extremely high. In particular, a solution with one parameter low and the other high (i.e., low-$f_{\mathrm{cov}}$/high-$N_{\mathrm{H}}$ or high-$f_{\mathrm{cov}}$/low-$N_{\mathrm{H}}$) cannot reproduce the observed locus.

\begin{figure*}
    \centering
    \includegraphics[width=1\linewidth]{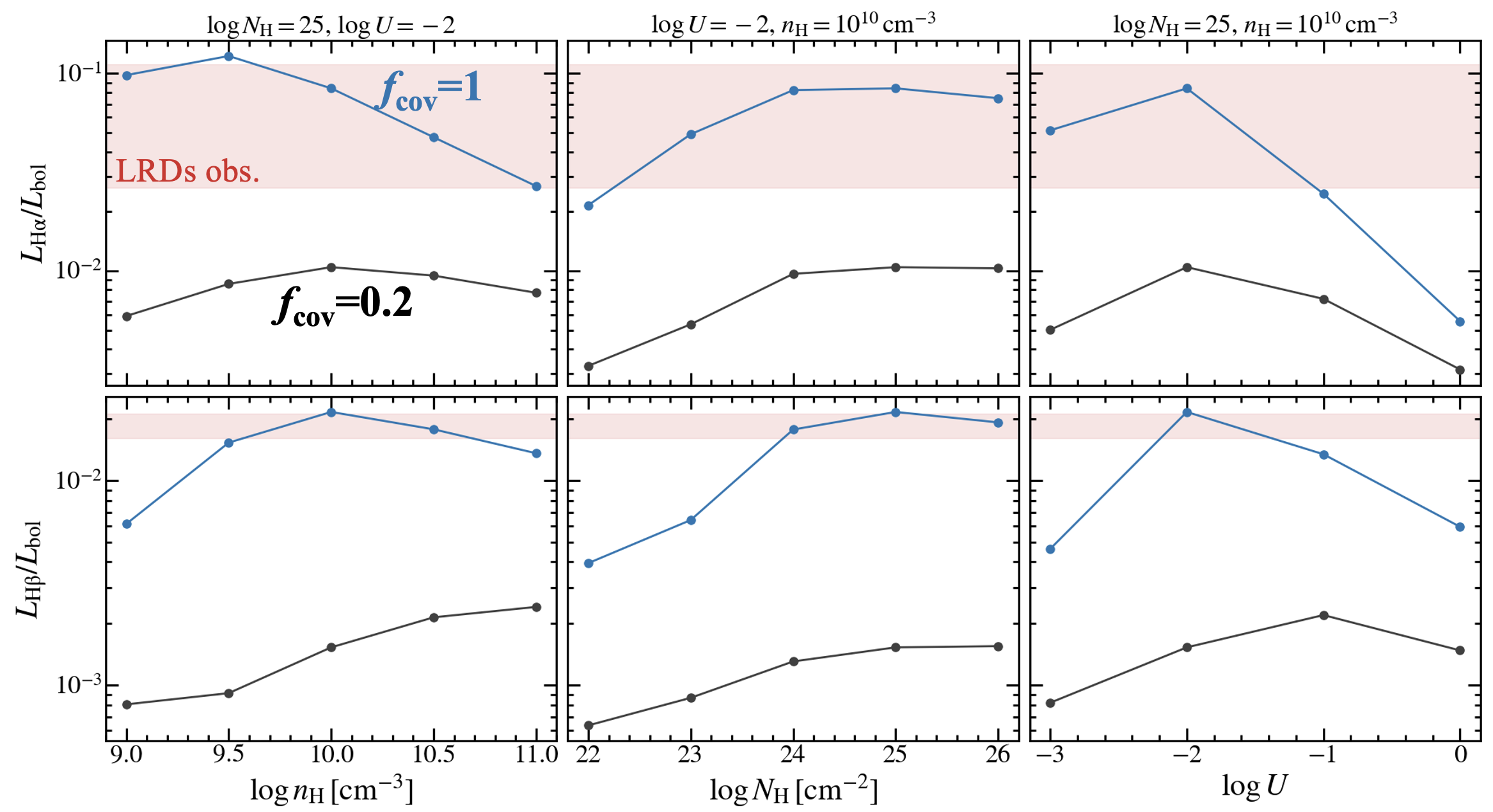}
    \caption{Broad Balmer line to bolometric luminosity ratios as functions of density (left), column density (center), and ionization parameter (right). The blue and black lines show the  \textsc{Cloudy} predictions of $f_\mathrm{cov}=1$ and 0.2, respectively. The red shades show the 1$\sigma$ ranges of the LRD sample in this work. }
    \label{fig:cloudy_loc}
\end{figure*}

\begin{figure}
    \centering
    \includegraphics[width=1\linewidth]{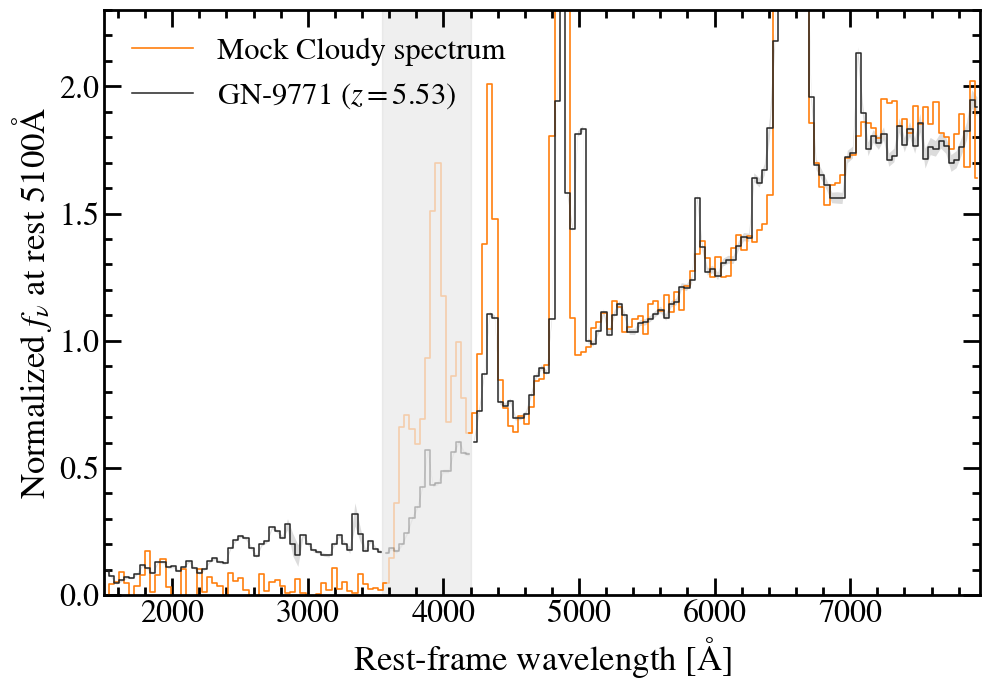}
    \caption{Mock JWST/NIRSpec PRISM spectrum generated from the \textsc{Cloudy} model. The orange line shows the simulated spectrum, 
    assuming $f_{\rm cov}=1$, $N_\mathrm{H}=10^{25}\,\mathrm{cm^{-2}}$, $n_\mathrm{H}=10^{10}\,\mathrm{cm^{-3}}$, and $\log U = -2$. The black line shows the observed PRISM spectrum of the LRD in our sample GN-9771. The gray shaded region indicate the region affected by the \textsc{Cloudy} artifact due to the finite number of hydrogen energy levels considered in the calculation \citep{Ji+2025, Naidu+2025}. These spectra are normalized at rest-frame 5100\AA\, and binned in three-pixel bins to highlight the overall continuum shape.}
    \label{fig:cloudy_mock_spectrum}
\end{figure}

\begin{figure*}
    \centering
    \includegraphics[width=1\linewidth]{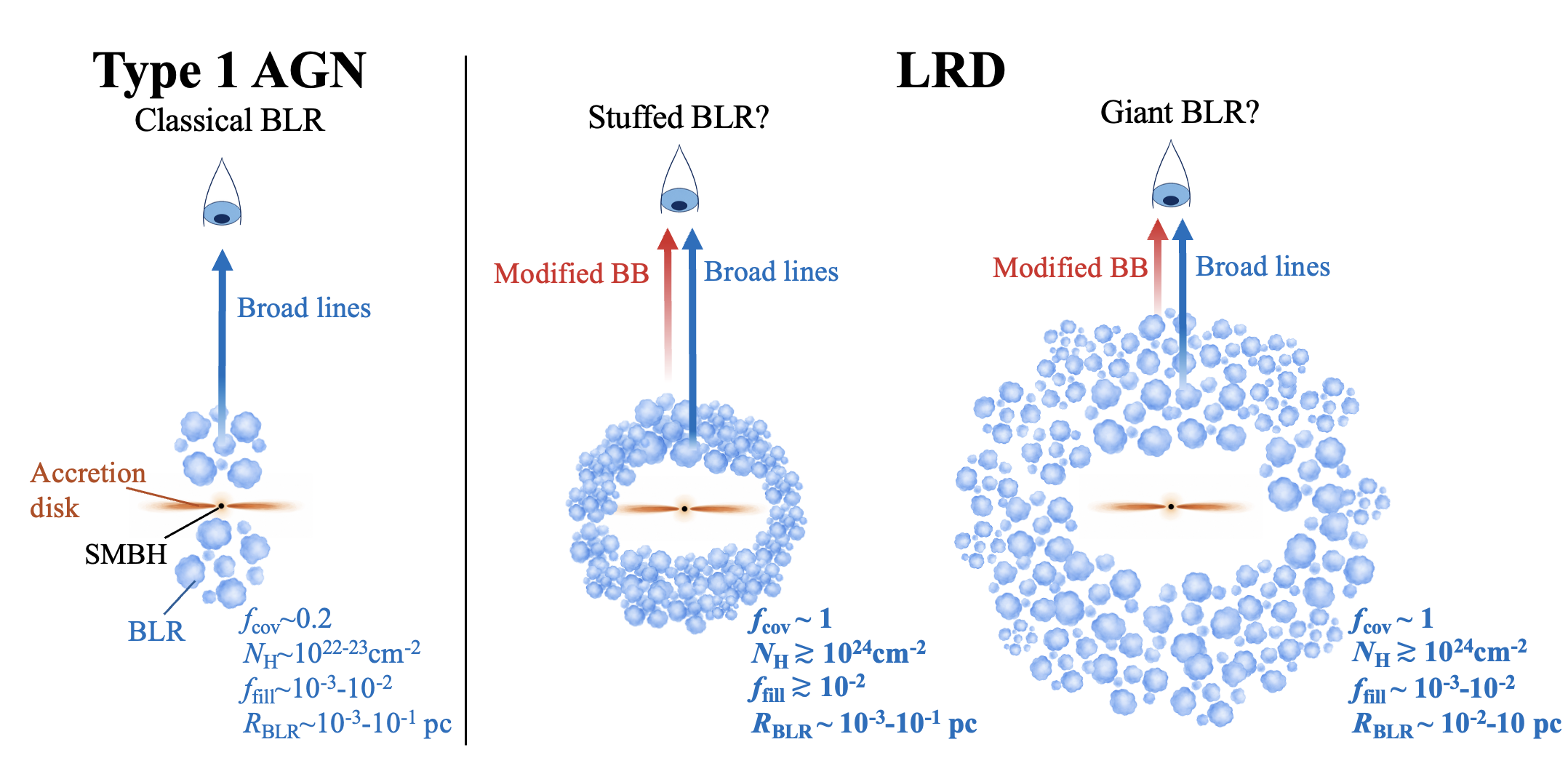}
    \caption{Schematic illustrations of the standard BLR (left), ``stuffed BLR" (center), and ``giant BLR" (right). The standard BLR typically have covering factor of $f_\mathrm{cov}\sim0.2$ \citep[e.g.,][]{Baldwin+1995,Korista_Goad_2004} and column density of $\log (N_\mathrm{H}/\mathrm{cm^{-2}}) = 22$--23 \citep[e.g.,][]{Peterson+2006}, while our \textsc{Cloudy} modeling suggests that the BLR of the LRDs have very high covering factor (almost unity) and large column density ($\log (N_\mathrm{H}/\mathrm{cm^{-2}}) \gtrsim 25$). The extremely large column density can be explained either by high filling factor (stuffed BLR) or by large BLR size (giant BLR). Either the stuffed BLR or giant BLR enhances the Balmer line emission (Figure \ref{fig:cloudy}) and produces the red optical continuum (Figure \ref{fig:cloudy_mock_spectrum}).}
    \label{fig:picture}
\end{figure*}

In Figure \ref{fig:cloudy_loc}, we demonstrate how Balmer line to bolometric luminosity ratios ($L_\mathrm{H\alpha,broad}/\lbol$ and $L_\mathrm{H\beta,broad}/\lbol$) depend on gas density (left), column density (center), and ionization parameter (right). We find that $L_\mathrm{H\alpha,broad}/\lbol$ and $L_\mathrm{H\beta,broad}/\lbol$ peaks at $\log (n_\mathrm{H}/\mathrm{cm^{-3}})\sim9.5$--10, $\log (N_\mathrm{H}/\mathrm{cm^{-2}})\sim24$--25, and $\log U\sim-2$. Together with $f_\mathrm{cov}\sim1$, these parameter values reproduce the observed $L_\mathrm{H\alpha,broad}/\lbol$ and $L_\mathrm{H\beta,broad}/\lbol$ values for LRDs. 
These $n_\mathrm{H}$ and $U$ values are typical for low-$z$ type 1 AGNs \citep[e.g.,][]{Davidson_Netzer_1979,Peterson+2006}, while the $N_\mathrm{H}$ value preferred for LRDs are one to two orders of magnitude higher than the typical values for low-$z$ type 1 AGNs \citep[$\sim10^{23}\,\mathrm{cm^{-2}}$; e.g.,][]{Peterson+2006}. However, the $f_\mathrm{cov}=0.2$ cases do not reproduce observed $L_\mathrm{H\alpha,broad}/\lbol$ and $L_\mathrm{H\beta,broad}/\lbol$ values for LRDs at any of $n_\mathrm{H}$, $N_\mathrm{H}$, and $U$ values, suggesting that the high covering factor is necessary. These results demonstrate that the gas density and ionization parameter of BLR clouds in the LRDs may be common to those of low-$z$ type 1 AGNs, while the large differences in the covering factor and column density enhance the Balmer line luminosities.

In Figure \ref{fig:cloudy_mock_spectrum}, we compare the mock observed spectrum predicted from our \textsc{Cloudy} model with one of the LRDs in our sample, GN-9771, as an exapmle. We find that the model with $f_{\mathrm{cov}}=1$ and $N_\mathrm{H}=10^{25}\,\mathrm{cm^{-2}}$ yields an optical continuum similar to that observed in LRDs (note that the excess in our \textsc{Cloudy} spectrum around the Balmer limit is \textsc{Cloudy} artifact due to the finite number of energy levels included in the calculation; \citealp{Ji+2025,Naidu+2025}). This demonstrates that our model can also reproduce the overall continuum shape of LRDs. 

We therefore conclude that the enhanced Balmer-line emission in LRDs can be explained, to first order, by a BLR with a high covering factor and a large gas column density. 
A BLR with such large $f_{\mathrm{cov}}$ and $N_{\mathrm{H}}$, while retaining standard $n_{\mathrm{H}}$ and $U$, suggests at least two possible scenarios (Figure~\ref{fig:picture}); one is a ``stuffed BLR,'' where the central engine is enshrouded by the BLR whose volume is filled by an extremely large number of clouds (i.e., a high volume filling factor $f_\mathrm{fill}$) within a radius comparable to that of normal BLRs. The other is a ``giant BLR,'' where the central engine is covered by the BLR whose radius is substantially larger, with a $f_\mathrm{fill}$ value similar to that of normal BLRs. Although the BLR filling factor for the normal AGNs has poorly been constrained, $f_\mathrm{fill}\sim10^{-3}$--$10^{-2}$ is typically assumed for BLRs \citep{Osterbrock_1991,Snedden_Gaskell_1999, Osterbrock_Ferland_2006}. The stuffed BLR scenario may represent a large filling factor of $f_\mathrm{fill}\gtrsim10^{-2}$. On the other hand, the giant BLR may have a size one or two orders of magnitude larger than that of a typical BLR \citep[$10^{-3}$--$10^{-1}$~pc; e.g.,][]{Peterson+2004,Kaspi+2000,Bentz+2013}. 
Note that these two scenarios should be regarded as limiting cases; the true BLR configuration may be somewhere between them.
These scenarios are in line with the emerging picture that LRDs host accreting black holes embedded in an optically thick, high-column-density gas cocoon \citep[e.g.,][]{Maiolino+2025,Inayoshi_Maiolino_2025,deGraaff+2025b,Naidu+2025,Rusakov+2026,Ando+2026,Sneppen+2026a}. 

The stuffed BLR or giant BLR scenarios may even explain the other observations of LRDs as follows:

\begin{itemize}
    \item Weak X-ray emission: the optically thick gas obscures the X-ray emission \citep[e.g.,][]{Ananna+2024,Yue+2024,Maiolino+2025,Sacchi+2025}.
    \item Balmer absorption lines: the optically thick gas with high density has a large column density of $n=2$ hydrogen atoms, which causes the Balmer absorption lines \citep[e.g.,][]{Matthee+2024,Kocevski+2025,DEugenio+2026}.
    \item Weak continuum variability of short timescale: 
    the continuum emission may at least partially be thermalized because of the large column density. In this case, it takes a long time for continuum photons to escape from the BLR and the high frequency component of continuum variability may be blurred \citep{Secunda+2026}, which is consistent with the weak short-term continuum variability observed in majority of LRDs \citep[e.g.,][]{Kokubo_Harikane_2025,ZZhang+2025a,Tee+2025,Stone+2025,Liu+2026,Burke+2025}. 
\end{itemize}
%
A full quantitative assessment of these points is beyond the scope of this paper and is left for future work.

\section{Summary}\label{sec:summary}
In this work, we use archival JWST/NIRSpec spectra and present statistical analysis of Balmer line and continuum luminosities of LRDs. Our analysis and results are summarized below:

\begin{itemize}
    \item We analyze JWST/NIRSpec PRISM and grating spectra of 37 little red dots (LRDs) at $z\sim3$--$7$. We decompose the H$\alpha$+[N\,\textsc{ii}] complex (and H$\beta$ where available) to isolate the broad components, and we estimate $L_{5100}$ and an empirical $L_{\rm bol}$ from modified blackbody continuum fits to the PRISM spectra.
    \item We identify a tight correlation between the broad H$\alpha$ luminosities and the continuum luminosities ($L_{5100}$ and $L_{\rm bol}$). LRDs are systematically offset from the type~1 AGN scaling relations, showing $\sim40$ times higher $\lbha$ at fixed $\lbol$, implying enhanced Balmer line emission efficiency. 
    \item We interpret this offset using \textsc{Cloudy} photoionization calculations within the classical LOC framework. The observed enhancement can be reproduced if LRDs have a BLR with a substantially high covering factor and large hydrogen column density than typical type~1 AGN, which simultaneously reproduce the modified blackbody continuum shape. Such gas conditions with high covering factor and large column density suggest either a very high volume filling factor or a large size of BLR, namely, ``stuffed BLR" or ``giant BLR," respectively.
\end{itemize}

\begin{acknowledgments}
We thank Yuki Isobe for providing the line-spread functions for JWST/NIRSpec. 
This work is based on observations made with the NASA/ ESA/CSA James Webb Space Telescope. The NIRSpec/IFU data were obtained from the Mikulski Archive for Space Telescopes at the Space Telescope Science Institute, which is operated by the Association of Universities for Research in Astronomy, Inc., under NASA contract NAS 5-03127 for JWST. The NIRSpec MSA data presented herein were retrieved from DJA. DJA is an initiative of the Cosmic Dawn Center (DAWN), which is funded by the Danish National Research Foundation under grant DNRF140. We thank DAWN for providing the reduced NIRSpec data. The observational data presented in this work are associated with programs 
ERS 1345 (CEERS; PI: Finkelstein), GTO 1180 (PI: Eisenstein), GTO 1181 (PI: Eisenstein), GTO 1210 (PI: Luetzgendorf), GTO 1286 (PI: Luetzgendorf), GO 4233 (PI: de Graaff), GO 4106 (PI: Nelson), GO 5664 (PI: Matthee), GO 5015 (PI: Übler and Maiolino), and GO 1287 (PI: Isaak). The authors acknowledge the teams conducting these observations for publicly releasing the data.
HY acknowledges support by KAKENHI (25KJ0832) through Japan Society for the Promotion of Science (JSPS). MO acknowledges the supports from the World Premier International Research Center Initiative (WPI Initiative), MEXT, Japan, the joint research program of the Institute for Cosmic Ray Research (ICRR), the University of Tokyo, and KAKENHI (21H04467, 25H00674) through Japan Society for the Promotion of Science (JSPS). YK acknowledges support by KAKENHI (26KJ0960) through JSPS, JSR Fellowship, and FoPM, WINGS Program, the University of Tokyo. TK acknowledges support by KAKENHI (26KJ1232) through Japan Society for the Promotion of Science (JSPS). MN is supported by JSPS KAKENHI Grant Nos. 25KJ0828. 
The authors acknowledge the use of ChatGPT (OpenAI, GPT-5.5) and Codex (OpenAI, Codex 5.4) to assist with language editing, code development, and figure generation. All AI-assisted outputs were carefully reviewed and validated by the authors. The authors take full responsibility for all analyses, interpretations, and conclusions presented in this work.

\end{acknowledgments}

\facilities{JWST}

\software{
grizli \citep{Brammer_2021, Brammer_2023}, emcee \citep{Foreman-Mackey+13}, Cloudy \citep{Gunasekera+2023}, 
Matplotlib \citep{Hunter_2007}, Astropy \citep{Astropy+2013,Astropy+2018,Astropy+2022}, NumPy \citep{Harris+2020} 
          }

\bibliography{ref}{}

@ARTICLE{Kiyota+2026_LRDdust,
       author = {{Kiyota}, Tomokazu and {Ouchi}, Masami and {Yanagisawa}, Hiroto and {Ando}, Makoto and {Harikane}, Yuichi and {Kageura}, Yuta and {Nakane}, Minami and {Nakazato}, Yurina and {Ono}, Yoshiaki and {Takeda}, Yui},
        title = "{ATLAS. III. Dust Around Little Red Dots: Hydrogen Line Ratios beyond Dust-free Non-Case B Models}",
      journal = {arXiv e-prints},
         year = 2026,
        month = aug,
          eid = {arXiv:2608.10832},
        pages = {arXiv:2608.10832},
          doi = {10.48550/arXiv.2608.10832},
archivePrefix = {arXiv},
       eprint = {2608.10832},
 primaryClass = {astro-ph.GA},
       adsurl = {https://ui.adsabs.harvard.edu/abs/2026arXiv260810832K}
}

@ARTICLE{Kageura+2026_LRDlya,
       author = {{Kageura}, Yuta and {Ouchi}, Masami and {Yanagisawa}, Hiroto and {Ando}, Makoto and {Harikane}, Yuichi and {Kiyota}, Tomokazu and {Nakane}, Minami and {Ono}, Yoshiaki and {Takeda}, Yui},
        title = "{ATLAS. IV. A JWST+MUSE Demographic Study of Ly$α$ Profiles in Little Red Dots}",
      journal = {arXiv e-prints},
         year = 2026,
        month = aug,
          eid = {arXiv:2608.14534},
        pages = {arXiv:2608.14534},
          doi = {10.48550/arXiv.2608.14534},
archivePrefix = {arXiv},
       eprint = {2608.14534},
 primaryClass = {astro-ph.GA},
       adsurl = {https://ui.adsabs.harvard.edu/abs/2026arXiv260814534K}
}

@ARTICLE{Yanagisawa+2026_balmerabs,
       author = {{Yanagisawa}, Hiroto and {Ouchi}, Masami and {Kiyota}, Tomokazu and {Ando}, Makoto and {Harikane}, Yuichi and {Kageura}, Yuta and {Nakane}, Minami and {Ono}, Yoshiaki and {Takeda}, Yui},
        title = "{ATLAS. II. Extremely High Incidence of Balmer Line Absorption with Predominant Blueshifts in LRDs: Statistical Insights through Comparison with Type 1 AGNs}",
      journal = {arXiv e-prints},
         year = 2026,
        month = jul,
          eid = {arXiv:2607.26269},
        pages = {arXiv:2607.26269},
          doi = {10.48550/arXiv.2607.26269},
archivePrefix = {arXiv},
       eprint = {2607.26269},
 primaryClass = {astro-ph.GA},
       adsurl = {https://ui.adsabs.harvard.edu/abs/2026arXiv260726269Y}
}

@ARTICLE{Torralba+2026,
       author = {{Torralba}, Alberto and {Matthee}, Jorryt and {Pezzulli}, Gabriele and {Naidu}, Rohan P. and {Ishikawa}, Yuzo and {Brammer}, Gabriel B. and {Chang}, Seok-Jun and {Chisholm}, John and {de Graaff}, Anna and {D'Eugenio}, Francesco and {Di Cesare}, Claudia and {Eilers}, Anna-Christina and {Greene}, Jenny E. and {Gronke}, Max and {Iani}, Edoardo and {Kokorev}, Vasily and {Kotiwale}, Gauri and {Kramarenko}, Ivan and {Ma}, Yilun and {Mascia}, Sara and {Navarrete}, Benjam{\'\i}n and {Nelson}, Erica and {Oesch}, Pascal and {Simcoe}, Robert A. and {Wuyts}, Stijn},
        title = "{The warm outer layer of a little red dot as the source of [Fe II] and collisional Balmer lines with scattering wings}",
      journal = {\aap},
         year = 2026,
        month = feb,
       volume = {707},
          eid = {A75},
        pages = {A75},
          doi = {10.1051/0004-6361/202557537},
archivePrefix = {arXiv},
       eprint = {2510.00103},
 primaryClass = {astro-ph.GA},
       adsurl = {https://ui.adsabs.harvard.edu/abs/2026A&A...707A..75T}
}

@ARTICLE{Shen_Liu_2012,
       author = {{Shen}, Yue and {Liu}, Xin},
        title = "{Comparing Single-epoch Virial Black Hole Mass Estimators for Luminous Quasars}",
      journal = {\apj},
         year = 2012,
        month = jul,
       volume = {753},
       number = {2},
          eid = {125},
        pages = {125},
          doi = {10.1088/0004-637X/753/2/125},
archivePrefix = {arXiv},
       eprint = {1203.0601},
 primaryClass = {astro-ph.CO},
       adsurl = {https://ui.adsabs.harvard.edu/abs/2012ApJ...753..125S}
}

@ARTICLE{Bentz+2013,
       author = {{Bentz}, Misty C. and {Denney}, Kelly D. and {Grier}, Catherine J. and {Barth}, Aaron J. and {Peterson}, Bradley M. and {Vestergaard}, Marianne and {Bennert}, Vardha N. and {Canalizo}, Gabriela and {De Rosa}, Gisella and {Filippenko}, Alexei V. and {Gates}, Elinor L. and {Greene}, Jenny E. and {Li}, Weidong and {Malkan}, Matthew A. and {Pogge}, Richard W. and {Stern}, Daniel and {Treu}, Tommaso and {Woo}, Jong-Hak},
        title = "{The Low-luminosity End of the Radius-Luminosity Relationship for Active Galactic Nuclei}",
      journal = {\apj},
         year = 2013,
        month = apr,
       volume = {767},
       number = {2},
          eid = {149},
        pages = {149},
          doi = {10.1088/0004-637X/767/2/149},
archivePrefix = {arXiv},
       eprint = {1303.1742},
 primaryClass = {astro-ph.CO},
       adsurl = {https://ui.adsabs.harvard.edu/abs/2013ApJ...767..149B}
}

@ARTICLE{Kaspi+2000,
       author = {{Kaspi}, Shai and {Smith}, Paul S. and {Netzer}, Hagai and {Maoz}, Dan and {Jannuzi}, Buell T. and {Giveon}, Uriel},
        title = "{Reverberation Measurements for 17 Quasars and the Size-Mass-Luminosity Relations in Active Galactic Nuclei}",
      journal = {\apj},
         year = 2000,
        month = apr,
       volume = {533},
       number = {2},
        pages = {631-649},
          doi = {10.1086/308704},
archivePrefix = {arXiv},
       eprint = {astro-ph/9911476},
 primaryClass = {astro-ph},
       adsurl = {https://ui.adsabs.harvard.edu/abs/2000ApJ...533..631K}
}

@ARTICLE{Peterson+2004,
       author = {{Peterson}, B.~M. and {Ferrarese}, L. and {Gilbert}, K.~M. and {Kaspi}, S. and {Malkan}, M.~A. and {Maoz}, D. and {Merritt}, D. and {Netzer}, H. and {Onken}, C.~A. and {Pogge}, R.~W. and {Vestergaard}, M. and {Wandel}, A.},
        title = "{Central Masses and Broad-Line Region Sizes of Active Galactic Nuclei. II. A Homogeneous Analysis of a Large Reverberation-Mapping Database}",
      journal = {\apj},
         year = 2004,
        month = oct,
       volume = {613},
       number = {2},
        pages = {682-699},
          doi = {10.1086/423269},
archivePrefix = {arXiv},
       eprint = {astro-ph/0407299},
 primaryClass = {astro-ph},
       adsurl = {https://ui.adsabs.harvard.edu/abs/2004ApJ...613..682P}
}

@BOOK{Osterbrock_Ferland_2006,
       author = {{Osterbrock}, Donald E. and {Ferland}, Gary J.},
        title = "{Astrophysics of gaseous nebulae and active galactic nuclei}",
         year = 2006,
       adsurl = {https://ui.adsabs.harvard.edu/abs/2006agna.book.....O}
}

@ARTICLE{Snedden_Gaskell_1999,
       author = {{Snedden}, Stephanie A. and {Gaskell}, C. Martin},
        title = "{The Effect of Abundance Variations on Estimates of the Densities of Broad-Line Region Clouds in Quasars}",
      journal = {\apjl},
         year = 1999,
        month = aug,
       volume = {521},
       number = {2},
        pages = {L91-L94},
          doi = {10.1086/312188},
       adsurl = {https://ui.adsabs.harvard.edu/abs/1999ApJ...521L..91S}
}

@ARTICLE{Osterbrock_1991,
       author = {{Osterbrock}, D.~E.},
        title = "{Active galactic nuclei}",
      journal = {Reports on Progress in Physics},
         year = 1991,
        month = apr,
       volume = {54},
       number = {4},
        pages = {579-633},
          doi = {10.1088/0034-4885/54/4/002},
       adsurl = {https://ui.adsabs.harvard.edu/abs/1991RPPh...54..579O}
}

@ARTICLE{DEugenio+2026,
       author = {{D'Eugenio}, Francesco and {Juod{\v{z}}balis}, Ignas and {Ji}, Xihan and {Scholtz}, Jan and {Maiolino}, Roberto and {Carniani}, Stefano and {Perna}, Michele and {Mazzolari}, Giovanni and {{\"U}bler}, Hannah and {Arribas}, Santiago and {Bhatawdekar}, Rachana and {Bunker}, Andrew J. and {Cresci}, Giovanni and {Curtis-Lake}, Emma and {Hainline}, Kevin and {Inayoshi}, Kohei and {Isobe}, Yuki and {Ji}, Zhiyuan and {Johnson}, Benjamin D. and {Jones}, Gareth C. and {Looser}, Tobias J. and {Nelson}, Erica J. and {Parlanti}, Eleonora and {Pusk{\'a}s}, D{\'a}vid and {Rinaldi}, Pierluigi and {Robertson}, Brant and {Rodr{\'\i}guez Del Pino}, Bruno and {Shivaei}, Irene and {Sun}, Fengwu and {Tacchella}, Sandro and {Venturi}, Giacomo and {Volonteri}, Marta and {Williams}, Christina C. and {Willmer}, Christopher N.~A. and {Willott}, Chris and {Witstok}, Joris},
        title = "{JADES and BlackTHUNDER: rest-frame Balmer-line absorption and the local environment in a Little Red Dot at z = 5}",
      journal = {\mnras},
         year = 2026,
        month = jan,
       volume = {545},
       number = {3},
          eid = {staf2117},
        pages = {staf2117},
          doi = {10.1093/mnras/staf2117},
archivePrefix = {arXiv},
       eprint = {2506.14870},
 primaryClass = {astro-ph.GA},
       adsurl = {https://ui.adsabs.harvard.edu/abs/2026MNRAS.545f2117D}
}

@ARTICLE{Sacchi+2025,
       author = {{Sacchi}, Andrea and {Bogd{\'a}n}, {\'A}kos},
        title = "{Chandra Rules Out Super-Eddington Accretion Models for Little Red Dots}",
      journal = {\apjl},
         year = 2025,
        month = aug,
       volume = {989},
       number = {2},
          eid = {L30},
        pages = {L30},
          doi = {10.3847/2041-8213/adf5c8},
archivePrefix = {arXiv},
       eprint = {2505.09669},
 primaryClass = {astro-ph.GA},
       adsurl = {https://ui.adsabs.harvard.edu/abs/2025ApJ...989L..30S}
}

@ARTICLE{Cloonan+2026,
       author = {{Cloonan}, Aidan P. and {Whitaker}, Katherine E. and {Manning}, Sinclaire M. and {Williams}, Christina C. and {Greene}, Jenny E. and {Oesch}, Pascal A. and {Weibel}, Andrea and {Brammer}, Gabriel and {de Graaff}, Anna and {Hviding}, Raphael E. and {Dayal}, Pratika and {Jespersen}, Christian Kragh and {Ji}, Zhiyuan and {Labbe}, Ivo and {Xiao}, Mengyuan and {Zhang}, Yunchong},
        title = "{A PANORAMIC of UV-optical morphologies of ``Little Red Dots'': Two groups of LRDs distinguished by UV half-light radius}",
      journal = {arXiv e-prints},
         year = 2026,
        month = mar,
          eid = {arXiv:2603.24700},
        pages = {arXiv:2603.24700},
          doi = {10.48550/arXiv.2603.24700},
archivePrefix = {arXiv},
       eprint = {2603.24700},
 primaryClass = {astro-ph.GA},
       adsurl = {https://ui.adsabs.harvard.edu/abs/2026arXiv260324700C}
}

@ARTICLE{Rinaldi+2026,
       author = {{Rinaldi}, Pierluigi and {Hainline}, Kevin and {D'Eugenio}, Francesco and {P{\'e}rez-Gonz{\'a}lez}, Pablo G. and {Eisenstein}, Daniel J. and {Willmer}, Christopher N.~A. and {Carreira}, Courtney and {Robertson}, Brant and {Johnson}, Benjamin D. and {Alberts}, Stacey and {Baker}, William M. and {Bunker}, Andrew J. and {Carniani}, Stefano and {Egami}, Eiichi and {Helton}, Jakob M. and {Ji}, Zhiyuan and {Juod{\v{z}}balis}, Ignas and {Lin}, Xiaojing and {Lyu}, Jianwei and {Ma}, Zheng and {Maiolino}, Roberto and {Parlanti}, Eleonora and {Scholtz}, Jan and {Sun}, Yang and {Tacchella}, Sandro and {Venturi}, Giacomo and {Williams}, Christina C. and {Willott}, Chris and {Witstok}, Joris and {Wu}, Zihao},
        title = "{The Way We Tally Becomes the Tale: the Impact of Selection Strategies on the Inferred Evolution of Little Red Dots Across Cosmic Time}",
      journal = {arXiv e-prints},
         year = 2026,
        month = apr,
          eid = {arXiv:2604.07138},
        pages = {arXiv:2604.07138},
          doi = {10.48550/arXiv.2604.07138},
archivePrefix = {arXiv},
       eprint = {2604.07138},
 primaryClass = {astro-ph.GA},
       adsurl = {https://ui.adsabs.harvard.edu/abs/2026arXiv260407138R}
}

@INCOLLECTION{Peterson+2006,
       author = {{Peterson}, B.~M.},
        title = "{The Broad-Line Region in Active Galactic Nuclei}",
    booktitle = {Physics of Active Galactic Nuclei at all Scales},
         year = 2006,
       editor = {{Alloin}, Danielle},
       volume = {693},
        pages = {77},
          doi = {10.1007/3-540-34621-X_3},
       adsurl = {https://ui.adsabs.harvard.edu/abs/2006LNP...693...77P}
}

@ARTICLE{Liu+2026,
       author = {{Liu}, Zhaoran and {Naidu}, Rohan P. and {Secunda}, Amy and {Greene}, Jenny E. and {Matthee}, Jorryt and {Chisholm}, John and {de Graaff}, Anna and {Robbins}, Luke and {Antwi-Danso}, Jacqueline and {Brammer}, Gabriel and {Sun}, Wendy Q. and {Eilers}, Anna-Christina and {Fujimoto}, Seiji and {Furtak}, Lukas J. and {Kara}, Erin and {Kokorev}, Vasily and {Marchesini}, Danilo and {Oesch}, Pascal A. and {Pierel}, Justin D.~R. and {Shen}, Xuejian and {Simcoe}, Robert A. and {Torralba}, Alberto and {Vogelsberger}, Mark},
        title = "{How I Wonder What You Are -- JWST's Little Red Dots do not TWINKLE}",
      journal = {arXiv e-prints},
         year = 2026,
        month = apr,
          eid = {arXiv:2604.13000},
        pages = {arXiv:2604.13000},
          doi = {10.48550/arXiv.2604.13000},
archivePrefix = {arXiv},
       eprint = {2604.13000},
 primaryClass = {astro-ph.GA},
       adsurl = {https://ui.adsabs.harvard.edu/abs/2026arXiv260413000L}
}

@ARTICLE{Secunda+2026,
       author = {{Secunda}, Amy and {Somerville}, Rachel S. and {Jiang}, Yan-Fei and {Greene}, Jenny E. and {Furtak}, Lukas J. and {Zitrin}, Adi},
        title = "{Do Little Red Dots Vary?}",
      journal = {\apj},
         year = 2026,
        month = jan,
       volume = {996},
       number = {1},
          eid = {6},
        pages = {6},
          doi = {10.3847/1538-4357/ae1f08},
archivePrefix = {arXiv},
       eprint = {2509.03571},
 primaryClass = {astro-ph.GA},
       adsurl = {https://ui.adsabs.harvard.edu/abs/2026ApJ...996....6S}
}

@ARTICLE{Davidson_Netzer_1979,
       author = {{Davidson}, Kris and {Netzer}, Hagai},
        title = "{The emission lines of quasars and similar objects}",
      journal = {Reviews of Modern Physics},
         year = 1979,
        month = oct,
       volume = {51},
       number = {4},
        pages = {715-766},
          doi = {10.1103/RevModPhys.51.715},
       adsurl = {https://ui.adsabs.harvard.edu/abs/1979RvMP...51..715D}
}

@ARTICLE{Korista+1997,
       author = {{Korista}, Kirk and {Baldwin}, Jack and {Ferland}, Gary and {Verner}, Dima},
        title = "{An Atlas of Computed Equivalent Widths of Quasar Broad Emission Lines}",
      journal = {\apjs},
         year = 1997,
        month = jan,
       volume = {108},
       number = {2},
        pages = {401-415},
          doi = {10.1086/312966},
archivePrefix = {arXiv},
       eprint = {astro-ph/9611220},
 primaryClass = {astro-ph},
       adsurl = {https://ui.adsabs.harvard.edu/abs/1997ApJS..108..401K}
}

@ARTICLE{Sneppen+2026b,
       author = {{Sneppen}, Albert and {Matthews}, James H. and {Watson}, Darach and {Cameron}, Alex J. and {Sim}, Stuart A. and {Witstok}, Joris and {Brammer}, Gabriel B. and {Heintz}, Kasper E. and {Nikopoulos}, Georgios},
        title = "{Paschen Jumps in Little Red Dots: Evidence for Nebular Continua}",
      journal = {arXiv e-prints},
         year = 2026,
        month = apr,
          eid = {arXiv:2604.09399},
        pages = {arXiv:2604.09399},
          doi = {10.48550/arXiv.2604.09399},
archivePrefix = {arXiv},
       eprint = {2604.09399},
 primaryClass = {astro-ph.GA},
       adsurl = {https://ui.adsabs.harvard.edu/abs/2026arXiv260409399S}
}

@ARTICLE{Sneppen+2026a,
       author = {{Sneppen}, A. and {Watson}, D. and {Matthews}, J.~H. and {Nikopoulos}, G. and {Allen}, N. and {Brammer}, G. and {Damgaard}, R. and {Heintz}, K.~E. and {Knigge}, C. and {Long}, K.~S. and {Rusakov}, V. and {Sim}, S.~A. and {Witstok}, J.},
        title = "{Inside the cocoon: a comprehensive explanation of the spectra of Little Red Dots}",
      journal = {arXiv e-prints},
         year = 2026,
        month = jan,
          eid = {arXiv:2601.18864},
        pages = {arXiv:2601.18864},
          doi = {10.48550/arXiv.2601.18864},
archivePrefix = {arXiv},
       eprint = {2601.18864},
 primaryClass = {astro-ph.GA},
       adsurl = {https://ui.adsabs.harvard.edu/abs/2026arXiv260118864S}
}

@ARTICLE{Pang+2026,
       author = {{Pang}, Yuxuan and {Wang}, Xin and {Cheng}, Cheng and {Wang}, Shengzhe and {Zhou}, Hang and {Zhou}, Qianqiao and {Wu}, Xue-Bing and {Glazebrook}, Karl},
        title = "{The Structure and Evolution of LRDs: Insights from JWST NIRSpec Medium and High Resolution Spectroscopy at $z\sim4$}",
      journal = {arXiv e-prints},
         year = 2026,
        month = feb,
          eid = {arXiv:2602.12548},
        pages = {arXiv:2602.12548},
          doi = {10.48550/arXiv.2602.12548},
archivePrefix = {arXiv},
       eprint = {2602.12548},
 primaryClass = {astro-ph.GA},
       adsurl = {https://ui.adsabs.harvard.edu/abs/2026arXiv260212548P}
}

@ARTICLE{Ando+2026,
       author = {{Ando}, Makoto and {Harikane}, Yuichi and {Katz}, Harley and {Inayoshi}, Kohei and {Tanaka}, Takumi S.},
        title = "{The UV Side of Little Red Dots: Red, Compact, and Iron-Enhanced Rest-UV Emission with a Strong Downturn around Ly$α$}",
      journal = {arXiv e-prints},
         year = 2026,
        month = jun,
          eid = {arXiv:2606.03522},
        pages = {arXiv:2606.03522},
archivePrefix = {arXiv},
       eprint = {2606.03522},
 primaryClass = {astro-ph.GA},
       adsurl = {https://ui.adsabs.harvard.edu/abs/2026arXiv260603522A}
}

@ARTICLE{Kiyota+2025,
       author = {{Kiyota}, Tomokazu and {Ouchi}, Masami and {Xu}, Yi and {Nakazato}, Yurina and {Soga}, Kenta and {Yajima}, Hidenobu and {Fujimoto}, Seiji and {Harikane}, Yuichi and {Nakajima}, Kimihiko and {Ono}, Yoshiaki and {Sun}, Dongsheng and {Kusakabe}, Haruka and {Ceverino}, Daniel and {Hatsukade}, Bunyo and {Iono}, Daisuke and {Kohno}, Kotaro and {Nakanishi}, Kouichiro},
        title = "{Comprehensive JWST+ALMA Study on the Extended Ly{\ensuremath{\alpha}} Emitters, Himiko, and CR7 at z {\ensuremath{\sim}} 7: Blue Major Merger Systems in Stark Contrast to Submillimeter Galaxies}",
      journal = {\apj},
         year = 2025,
        month = dec,
       volume = {995},
       number = {2},
          eid = {150},
        pages = {150},
          doi = {10.3847/1538-4357/ae1cc3},
archivePrefix = {arXiv},
       eprint = {2504.03156},
 primaryClass = {astro-ph.GA},
       adsurl = {https://ui.adsabs.harvard.edu/abs/2025ApJ...995..150K}
}

@ARTICLE{Valentino+2023,
       author = {{Valentino}, Francesco and {Brammer}, Gabriel and {Gould}, Katriona M.~L. and {Kokorev}, Vasily and {Fujimoto}, Seiji and {Jespersen}, Christian Kragh and {Vijayan}, Aswin P. and {Weaver}, John R. and {Ito}, Kei and {Tanaka}, Masayuki and {Ilbert}, Olivier and {Magdis}, Georgios E. and {Whitaker}, Katherine E. and {Faisst}, Andreas L. and {Gallazzi}, Anna and {Gillman}, Steven and {Gim{\'e}nez-Arteaga}, Clara and {G{\'o}mez-Guijarro}, Carlos and {Kubo}, Mariko and {Heintz}, Kasper E. and {Hirschmann}, Michaela and {Oesch}, Pascal and {Onodera}, Masato and {Rizzo}, Francesca and {Lee}, Minju and {Strait}, Victoria and {Toft}, Sune},
        title = "{An Atlas of Color-selected Quiescent Galaxies at z > 3 in Public JWST Fields}",
      journal = {\apj},
         year = 2023,
        month = apr,
       volume = {947},
       number = {1},
          eid = {20},
        pages = {20},
          doi = {10.3847/1538-4357/acbefa},
archivePrefix = {arXiv},
       eprint = {2302.10936},
 primaryClass = {astro-ph.GA},
       adsurl = {https://ui.adsabs.harvard.edu/abs/2023ApJ...947...20V}
}

@ARTICLE{Weaver+2024,
       author = {{Weaver}, John R. and {Cutler}, Sam E. and {Pan}, Richard and {Whitaker}, Katherine E. and {Labb{\'e}}, Ivo and {Price}, Sedona H. and {Bezanson}, Rachel and {Brammer}, Gabriel and {Marchesini}, Danilo and {Leja}, Joel and {Wang}, Bingjie and {Furtak}, Lukas J. and {Zitrin}, Adi and {Atek}, Hakim and {Chemerynska}, Iryna and {Coe}, Dan and {Dayal}, Pratika and {van Dokkum}, Pieter and {Feldmann}, Robert and {F{\"o}rster Schreiber}, Natascha M. and {Franx}, Marijn and {Fujimoto}, Seiji and {Fudamoto}, Yoshinobu and {Glazebrook}, Karl and {de Graaff}, Anna and {Greene}, Jenny E. and {Juneau}, St{\'e}phanie and {Kassin}, Susan and {Kriek}, Mariska and {Khullar}, Gourav and {Maseda}, Michael V. and {Mowla}, Lamiya A. and {Muzzin}, Adam and {Nanayakkara}, Themiya and {Nelson}, Erica J. and {Oesch}, Pascal A. and {Pacifici}, Camilla and {Papovich}, Casey and {Setton}, David J. and {Shapley}, Alice E. and {Shipley}, Heath V. and {Smit}, Renske and {Stefanon}, Mauro and {Taylor}, Edward N. and {Weibel}, Andrea and {Williams}, Christina C.},
        title = "{The UNCOVER Survey: A First-look HST + JWST Catalog of 60,000 Galaxies near A2744 and beyond}",
      journal = {\apjs},
         year = 2024,
        month = jan,
       volume = {270},
       number = {1},
          eid = {7},
        pages = {7},
          doi = {10.3847/1538-4365/ad07e0},
archivePrefix = {arXiv},
       eprint = {2301.02671},
 primaryClass = {astro-ph.GA},
       adsurl = {https://ui.adsabs.harvard.edu/abs/2024ApJS..270....7W}
}

@ARTICLE{Suess+2024,
       author = {{Suess}, Katherine A. and {Weaver}, John R. and {Price}, Sedona H. and {Pan}, Richard and {Wang}, Bingjie and {Bezanson}, Rachel and {Brammer}, Gabriel and {Cutler}, Sam E. and {Labb{\'e}}, Ivo and {Leja}, Joel and {Williams}, Christina C. and {Whitaker}, Katherine E. and {Atek}, Hakim and {Dayal}, Pratika and {de Graaff}, Anna and {Feldmann}, Robert and {Franx}, Marijn and {Fudamoto}, Yoshinobu and {Fujimoto}, Seiji and {Furtak}, Lukas J. and {Goulding}, Andy D. and {Greene}, Jenny E. and {Khullar}, Gourav and {Kokorev}, Vasily and {Kriek}, Mariska and {Lorenz}, Brian and {Marchesini}, Danilo and {Maseda}, Michael V. and {Matthee}, Jorryt and {Miller}, Tim B. and {Mitsuhashi}, Ikki and {Mowla}, Lamiya A. and {Muzzin}, Adam and {Naidu}, Rohan P. and {Nanayakkara}, Themiya and {Nelson}, Erica J. and {Oesch}, Pascal A. and {Setton}, David J. and {Shipley}, Heath and {Smit}, Renske and {Spilker}, Justin S. and {van Dokkum}, Pieter and {Zitrin}, Adi},
        title = "{Medium Bands, Mega Science: A JWST/NIRCam Medium-band Imaging Survey of A2744}",
      journal = {\apj},
         year = 2024,
        month = nov,
       volume = {976},
       number = {1},
          eid = {101},
        pages = {101},
          doi = {10.3847/1538-4357/ad75fe},
archivePrefix = {arXiv},
       eprint = {2404.13132},
 primaryClass = {astro-ph.GA},
       adsurl = {https://ui.adsabs.harvard.edu/abs/2024ApJ...976..101S}
}

@ARTICLE{Price+2025,
       author = {{Price}, Sedona H. and {Bezanson}, Rachel and {Labbe}, Ivo and {Furtak}, Lukas J. and {de Graaff}, Anna and {Greene}, Jenny E. and {Kokorev}, Vasily and {Setton}, David J. and {Suess}, Katherine A. and {Brammer}, Gabriel and {Cutler}, Sam E. and {Leja}, Joel and {Pan}, Richard and {Wang}, Bingjie and {Weaver}, John R. and {Whitaker}, Katherine E. and {Atek}, Hakim and {Burgasser}, Adam J. and {Chemerynska}, Iryna and {Dayal}, Pratika and {Feldmann}, Robert and {F{\"o}rster Schreiber}, Natascha M. and {Fudamoto}, Yoshinobu and {Fujimoto}, Seiji and {Glazebrook}, Karl and {Goulding}, Andy D. and {Khullar}, Gourav and {Kriek}, Mariska and {Marchesini}, Danilo and {Maseda}, Michael V. and {Miller}, Tim B. and {Muzzin}, Adam and {Nanayakkara}, Themiya and {Nelson}, Erica and {Oesch}, Pascal A. and {Shipley}, Heath and {Smit}, Renske and {Taylor}, Edward N. and {Dokkum}, Pieter van and {Williams}, Christina C. and {Zitrin}, Adi},
        title = "{The UNCOVER Survey: First Release of Ultradeep JWST/NIRSpec PRISM Spectra for {\ensuremath{\sim}}700 Galaxies from z {\ensuremath{\sim}} 0.3─13 in A2744}",
      journal = {\apj},
         year = 2025,
        month = mar,
       volume = {982},
       number = {1},
          eid = {51},
        pages = {51},
          doi = {10.3847/1538-4357/adaec1},
archivePrefix = {arXiv},
       eprint = {2408.03920},
 primaryClass = {astro-ph.GA},
       adsurl = {https://ui.adsabs.harvard.edu/abs/2025ApJ...982...51P}
}

@ARTICLE{Maiolino+2025,
       author = {{Maiolino}, Roberto and {Risaliti}, Guido and {Signorini}, Matilde and {Trefoloni}, Bartolomeo and {Juod{\v{z}}balis}, Ignas and {Scholtz}, Jan and {{\"U}bler}, Hannah and {D'Eugenio}, Francesco and {Carniani}, Stefano and {Fabian}, Andy and {Ji}, Xihan and {Mazzolari}, Giovanni and {Bertola}, Elena and {Brusa}, Marcella and {Bunker}, Andrew J. and {Charlot}, Stephane and {Comastri}, Andrea and {Cresci}, Giovanni and {DeCoursey}, Christa Noel and {Egami}, Eiichi and {Fiore}, Fabrizio and {Gilli}, Roberto and {Perna}, Michele and {Tacchella}, Sandro and {Venturi}, Giacomo},
        title = "{JWST meets Chandra: a large population of Compton thick, feedback-free, and intrinsically X-ray weak AGN, with a sprinkle of SNe}",
      journal = {\mnras},
         year = 2025,
        month = apr,
       volume = {538},
       number = {3},
        pages = {1921-1943},
          doi = {10.1093/mnras/staf359},
archivePrefix = {arXiv},
       eprint = {2405.00504},
 primaryClass = {astro-ph.GA},
       adsurl = {https://ui.adsabs.harvard.edu/abs/2025MNRAS.538.1921M}
}

@ARTICLE{Wu_Shen_2022,
       author = {{Wu}, Qiaoya and {Shen}, Yue},
        title = "{A Catalog of Quasar Properties from Sloan Digital Sky Survey Data Release 16}",
      journal = {\apjs},
         year = 2022,
        month = dec,
       volume = {263},
       number = {2},
          eid = {42},
        pages = {42},
          doi = {10.3847/1538-4365/ac9ead},
archivePrefix = {arXiv},
       eprint = {2209.03987},
 primaryClass = {astro-ph.GA},
       adsurl = {https://ui.adsabs.harvard.edu/abs/2022ApJS..263...42W}
}

@ARTICLE{XLin+2026,
       author = {{Lin}, Xiaojing and {Fan}, Xiaohui and {Cai}, Zheng and {Liu}, Yichen and {Sun}, Fengwu and {Bian}, Fuyan and {Li}, Mingyu and {Mao}, Junjie and {Greene}, Jenny E. and {Liu}, Hanpu and {Li}, Jiaxuan and {Liu}, Weizhe and {Ma}, Yilun and {Sun}, Zechang and {Zhang}, Zijian},
        title = "{(LRDs)$^2$: The Low-ReDshift Little Red Dots Survey. II. DESI DR1 Sample}",
      journal = {arXiv e-prints},
         year = 2026,
        month = may,
          eid = {arXiv:2605.21574},
        pages = {arXiv:2605.21574},
          doi = {10.48550/arXiv.2605.21574},
archivePrefix = {arXiv},
       eprint = {2605.21574},
 primaryClass = {astro-ph.GA},
       adsurl = {https://ui.adsabs.harvard.edu/abs/2026arXiv260521574L}
}

@ARTICLE{Galavis+1997,
       author = {{Galavis}, M.~E. and {Mendoza}, C. and {Zeippen}, C.~J.},
        title = "{Atomic data from the IRON Project. XXII. Radiative rates for forbidden transitions within the ground configuration of ions in the carbon and oxygen isoelectronic sequences}",
      journal = {\aaps},
         year = 1997,
        month = may,
       volume = {123},
        pages = {159-171},
          doi = {10.1051/aas:1997344},
       adsurl = {https://ui.adsabs.harvard.edu/abs/1997A&AS..123..159G}
}

@ARTICLE{Greene_Ho_2005,
       author = {{Greene}, Jenny E. and {Ho}, Luis C.},
        title = "{Estimating Black Hole Masses in Active Galaxies Using the H{\ensuremath{\alpha}} Emission Line}",
      journal = {\apj},
         year = 2005,
        month = sep,
       volume = {630},
       number = {1},
        pages = {122-129},
          doi = {10.1086/431897},
archivePrefix = {arXiv},
       eprint = {astro-ph/0508335},
 primaryClass = {astro-ph},
       adsurl = {https://ui.adsabs.harvard.edu/abs/2005ApJ...630..122G}
}

@ARTICLE{Setton+2025,
       author = {{Setton}, David J. and {Greene}, Jenny E. and {Spilker}, Justin S. and {Williams}, Christina C. and {Labb{\'e}}, Ivo and {Ma}, Yilun and {Wang}, Bingjie and {Whitaker}, Katherine E. and {Leja}, Joel and {de Graaff}, Anna and {Alberts}, Stacey and {Bezanson}, Rachel and {Boogaard}, Leindert A. and {Brammer}, Gabriel and {Cutler}, Sam E. and {Cleri}, Nikko J. and {Cooper}, Olivia R. and {Dayal}, Pratika and {Fujimoto}, Seiji and {Furtak}, Lukas J. and {Goulding}, Andy D. and {Hirschmann}, Michaela and {Kokorev}, Vasily and {Maseda}, Michael V. and {McConachie}, Ian and {Matthee}, Jorryt and {Miller}, Tim B. and {Naidu}, Rohan P. and {Oesch}, Pascal A. and {Pan}, Richard and {Price}, Sedona H. and {Suess}, Katherine A. and {Weaver}, John R. and {Xiao}, Mengyuan and {Zhang}, Yunchong and {Zitrin}, Adi},
        title = "{A Confirmed Deficit of Hot and Cold Dust Emission in the Most Luminous Little Red Dots}",
      journal = {\apjl},
         year = 2025,
        month = sep,
       volume = {991},
       number = {1},
          eid = {L10},
        pages = {L10},
          doi = {10.3847/2041-8213/ade78b},
archivePrefix = {arXiv},
       eprint = {2503.02059},
 primaryClass = {astro-ph.GA},
       adsurl = {https://ui.adsabs.harvard.edu/abs/2025ApJ...991L..10S}
}

@ARTICLE{Labbe+2024,
       author = {{Labbe}, Ivo and {Greene}, Jenny E. and {Matthee}, Jorryt and {Treiber}, Helena and {Kokorev}, Vasily and {Miller}, Tim B. and {Kramarenko}, Ivan and {Setton}, David J. and {Ma}, Yilun and {Goulding}, Andy D. and {Bezanson}, Rachel and {Naidu}, Rohan P. and {Williams}, Christina C. and {Atek}, Hakim and {Brammer}, Gabriel and {Cutler}, Sam E. and {Chemerynska}, Iryna and {Cloonan}, Aidan P. and {Dayal}, Pratika and {de Graaff}, Anna and {Fudamoto}, Yoshinobu and {Fujimoto}, Seiji and {Furtak}, Lukas J. and {Glazebrook}, Karl and {Heintz}, Kasper E. and {Leja}, Joel and {Marchesini}, Danilo and {Nanayakkara}, Themiya and {Nelson}, Erica J. and {Oesch}, Pascal A. and {Pan}, Richard and {Price}, Sedona H. and {Shivaei}, Irene and {Sobral}, David and {Suess}, Katherine A. and {van Dokkum}, Pieter and {Wang}, Bingjie and {Weaver}, John R. and {Whitaker}, Katherine E. and {Zitrin}, Adi},
        title = "{An unambiguous AGN and a Balmer break in an Ultraluminous Little Red Dot at z=4.47 from Ultradeep UNCOVER and All the Little Things Spectroscopy}",
      journal = {arXiv e-prints},
         year = 2024,
        month = dec,
          eid = {arXiv:2412.04557},
        pages = {arXiv:2412.04557},
          doi = {10.48550/arXiv.2412.04557},
archivePrefix = {arXiv},
       eprint = {2412.04557},
 primaryClass = {astro-ph.GA},
       adsurl = {https://ui.adsabs.harvard.edu/abs/2024arXiv241204557L}
}

@ARTICLE{Isobe+2023,
       author = {{Isobe}, Yuki and {Ouchi}, Masami and {Nakajima}, Kimihiko and {Harikane}, Yuichi and {Ono}, Yoshiaki and {Xu}, Yi and {Zhang}, Yechi and {Umeda}, Hiroya},
        title = "{Redshift Evolution of Electron Density in the Interstellar Medium at z   0-9 Uncovered with JWST/NIRSpec Spectra and Line-spread Function Determinations}",
      journal = {\apj},
         year = 2023,
        month = oct,
       volume = {956},
       number = {2},
          eid = {139},
        pages = {139},
          doi = {10.3847/1538-4357/acf376},
archivePrefix = {arXiv},
       eprint = {2301.06811},
 primaryClass = {astro-ph.GA},
       adsurl = {https://ui.adsabs.harvard.edu/abs/2023ApJ...956..139I}
}

@ARTICLE{Baldwin+1995,
       author = {{Baldwin}, Jack and {Ferland}, Gary and {Korista}, Kirk and {Verner}, Dima},
        title = "{Locally Optimally Emitting Clouds and the Origin of Quasar Emission Lines}",
      journal = {\apjl},
         year = 1995,
        month = dec,
       volume = {455},
        pages = {L119},
          doi = {10.1086/309827},
archivePrefix = {arXiv},
       eprint = {astro-ph/9510080},
 primaryClass = {astro-ph},
       adsurl = {https://ui.adsabs.harvard.edu/abs/1995ApJ...455L.119B}
}

@ARTICLE{Korista_Goad_2004,
       author = {{Korista}, Kirk T. and {Goad}, Michael R.},
        title = "{What the Optical Recombination Lines Can Tell Us about the Broad-Line Regions of Active Galactic Nuclei}",
      journal = {\apj},
         year = 2004,
        month = may,
       volume = {606},
       number = {2},
        pages = {749-762},
          doi = {10.1086/383193},
archivePrefix = {arXiv},
       eprint = {astro-ph/0402506},
 primaryClass = {astro-ph},
       adsurl = {https://ui.adsabs.harvard.edu/abs/2004ApJ...606..749K}
}

@ARTICLE{Foreman-Mackey+13,
       author = {{Foreman-Mackey}, Daniel and {Hogg}, David W. and {Lang}, Dustin and {Goodman}, Jonathan},
        title = "{emcee: The MCMC Hammer}",
      journal = {\pasp},
         year = 2013,
        month = mar,
       volume = {125},
       number = {925},
        pages = {306},
          doi = {10.1086/670067},
archivePrefix = {arXiv},
       eprint = {1202.3665},
 primaryClass = {astro-ph.IM},
       adsurl = {https://ui.adsabs.harvard.edu/abs/2013PASP..125..306F}
}

@ARTICLE{Valentino+2025,
       author = {{Valentino}, F. and {Heintz}, K.~E. and {Brammer}, G. and {Ito}, K. and {Kokorev}, V. and {Whitaker}, K.~E. and {Gallazzi}, A. and {de Graaff}, A. and {Weibel}, A. and {Frye}, B.~L. and {Kamieneski}, P.~S. and {Jin}, S. and {Ceverino}, D. and {Faisst}, A. and {Farcy}, M. and {Fujimoto}, S. and {Gillman}, S. and {Gottumukkala}, R. and {Hamadouche}, M. and {Harrington}, K.~C. and {Hirschmann}, M. and {Jespersen}, C.~K. and {Kakimoto}, T. and {Kubo}, M. and {Lagos}, C. d. P. and {Lee}, M. and {Magdis}, G.~E. and {Man}, A.~W.~S. and {Onodera}, M. and {Rizzo}, F. and {Shimakawa}, R. and {Setton}, D.~J. and {Tanaka}, M. and {Toft}, S. and {Wu}, P.-F. and {Zhu}, P.},
        title = "{Gas outflows in two recently quenched galaxies at z = 4 and 7}",
      journal = {\aap},
         year = 2025,
        month = jul,
       volume = {699},
          eid = {A358},
        pages = {A358},
          doi = {10.1051/0004-6361/202553908},
archivePrefix = {arXiv},
       eprint = {2503.01990},
 primaryClass = {astro-ph.GA},
       adsurl = {https://ui.adsabs.harvard.edu/abs/2025A&A...699A.358V}
}

@ARTICLE{Heintz+2024,
       author = {{Heintz}, Kasper E. and {Watson}, Darach and {Brammer}, Gabriel and {Vejlgaard}, Simone and {Hutter}, Anne and {Strait}, Victoria B. and {Matthee}, Jorryt and {Oesch}, Pascal A. and {Jakobsson}, P{\'a}ll and {Tanvir}, Nial R. and {Laursen}, Peter and {Naidu}, Rohan P. and {Mason}, Charlotte A. and {Killi}, Meghana and {Jung}, Intae and {Hsiao}, Tiger Yu-Yang and {Abdurro'uf} and {Coe}, Dan and {Arrabal Haro}, Pablo and {Finkelstein}, Steven L. and {Toft}, Sune},
        title = "{Strong damped Lyman-{\ensuremath{\alpha}} absorption in young star-forming galaxies at redshifts 9 to 11}",
      journal = {Science},
         year = 2024,
        month = may,
       volume = {384},
       number = {6698},
        pages = {890-894},
          doi = {10.1126/science.adj0343},
archivePrefix = {arXiv},
       eprint = {2306.00647},
 primaryClass = {astro-ph.GA},
       adsurl = {https://ui.adsabs.harvard.edu/abs/2024Sci...384..890H}
}

@ARTICLE{deGraaff+2025a,
       author = {{de Graaff}, Anna and {Brammer}, Gabriel and {Weibel}, Andrea and {Lewis}, Zach and {Maseda}, Michael V. and {Oesch}, Pascal A. and {Bezanson}, Rachel and {Boogaard}, Leindert A. and {Cleri}, Nikko J. and {Cooper}, Olivia R. and {Gottumukkala}, Rashmi and {Greene}, Jenny E. and {Hirschmann}, Michaela and {Hviding}, Raphael E. and {Katz}, Harley and {Labb{\'e}}, Ivo and {Leja}, Joel and {Matthee}, Jorryt and {McConachie}, Ian and {Miller}, Tim B. and {Naidu}, Rohan P. and {Price}, Sedona H. and {Rix}, Hans-Walter and {Setton}, David J. and {Suess}, Katherine A. and {Wang}, Bingjie and {Whitaker}, Katherine E. and {Williams}, Christina C.},
        title = "{RUBIES: A complete census of the bright and red distant Universe with JWST/NIRSpec}",
      journal = {\aap},
         year = 2025,
        month = may,
       volume = {697},
          eid = {A189},
        pages = {A189},
          doi = {10.1051/0004-6361/202452186},
archivePrefix = {arXiv},
       eprint = {2409.05948},
 primaryClass = {astro-ph.GA},
       adsurl = {https://ui.adsabs.harvard.edu/abs/2025A&A...697A.189D}
}

@ARTICLE{Scholtz+2026,
       author = {{Scholtz}, J. and {D'Eugenio}, F. and {Maiolino}, R. and {Brazzini}, M. and {{\"U}bler}, H. and {Ji}, X. and {Perna}, M. and {Sun}, F. and {Brocchi}, G. and {Carniani}, S. and {Cresci}, G. and {Ivey}, L.~R. and {Juod{\v{z}}balis}, I. and {Marconi}, A. and {Mazzolari}, G. and {Risaliti}, G. and {Trefoloni}, B.},
        title = "{Little Red and Blue Dots: simply stratified Broad Line Regions}",
      journal = {arXiv e-prints},
         year = 2026,
        month = mar,
          eid = {arXiv:2603.22277},
        pages = {arXiv:2603.22277},
          doi = {10.48550/arXiv.2603.22277},
archivePrefix = {arXiv},
       eprint = {2603.22277},
 primaryClass = {astro-ph.GA},
       adsurl = {https://ui.adsabs.harvard.edu/abs/2026arXiv260322277S}
}

@ARTICLE{Matthee+2026,
       author = {{Matthee}, Jorryt and {Torralba}, Alberto and {Pezzulli}, Gabriele and {Naidu}, Rohan P. and {Chisholm}, John and {Mascia}, Sara and {Greene}, Jenny E. and {Ishikawa}, Yuzo and {Gronke}, Max and {Wuyts}, Stijn and {Bordoloi}, Rongmon and {Brammer}, Gabriel and {Chang}, Seok-Jun and {Eilers}, Anna-Christina and {de Graaff}, Anna and {Hviding}, Raphael E. and {Iani}, Edoardo and {Illingworth}, Garth and {Kashino}, Daichi and {Labbe}, Ivo and {Ma}, Yilun and {Maseda}, Michael V. and {Meyer}, Romain and {Nelson}, Erica and {Oesch}, Pascal and {Xiao}, Mengyuan},
        title = "{The Engine and its Flows: Little Red Dot spectra are shaped by the column densities of their gas envelopes}",
      journal = {arXiv e-prints},
         year = 2026,
        month = mar,
          eid = {arXiv:2603.17667},
        pages = {arXiv:2603.17667},
          doi = {10.48550/arXiv.2603.17667},
archivePrefix = {arXiv},
       eprint = {2603.17667},
 primaryClass = {astro-ph.GA},
       adsurl = {https://ui.adsabs.harvard.edu/abs/2026arXiv260317667M}
}

@ARTICLE{Greene+2025,
       author = {{Greene}, Jenny E. and {Setton}, David J. and {Furtak}, Lukas J. and {Naidu}, Rohan P. and {Volonteri}, Marta and {Dayal}, Pratika and {Labbe}, Ivo and {van Dokkum}, Pieter and {Bezanson}, Rachel and {Brammer}, Gabriel and {Cutler}, Sam E. and {Glazebrook}, Karl and {de Graaff}, Anna and {Hirschmann}, Michaela and {Hviding}, Raphael E. and {Kokorev}, Vasily and {Leja}, Joel and {Liu}, Hanpu and {Ma}, Yilun and {Matthee}, Jorryt and {Nanayakkara}, Themiya and {Oesch}, Pascal A. and {Pan}, Richard and {Price}, Sedona H. and {Spilker}, Justin S. and {Wang}, Bingjie and {Weaver}, John R. and {Whitaker}, Katherine E. and {Williams}, Christina C. and {Zitrin}, Adi},
        title = "{What you see is what you get: empirically measured bolometric luminosities of Little Red Dots}",
      journal = {arXiv e-prints},
         year = 2025,
        month = sep,
          eid = {arXiv:2509.05434},
        pages = {arXiv:2509.05434},
          doi = {10.48550/arXiv.2509.05434},
archivePrefix = {arXiv},
       eprint = {2509.05434},
 primaryClass = {astro-ph.GA},
       adsurl = {https://ui.adsabs.harvard.edu/abs/2025arXiv250905434G}
}

@ARTICLE{ZZhang+2025a,
       author = {{Zhang}, Zijian and {Jiang}, Linhua and {Liu}, Weiyang and {Ho}, Luis C.},
        title = "{Analysis of Multi-epoch JWST Images of {\ensuremath{\sim}}300 Little Red Dots: Tentative Detection of Variability in a Minority of Sources}",
      journal = {\apj},
         year = 2025,
        month = may,
       volume = {985},
       number = {1},
          eid = {119},
        pages = {119},
          doi = {10.3847/1538-4357/adcb3e},
archivePrefix = {arXiv},
       eprint = {2411.02729},
 primaryClass = {astro-ph.GA},
       adsurl = {https://ui.adsabs.harvard.edu/abs/2025ApJ...985..119Z}
}

@ARTICLE{Stone+2025,
       author = {{Stone}, Zachary and {Shen}, Yue and {Zhuang}, Ming-Yang and {Hu}, Lei and {Pierel}, Justin and {Li}, Junyao and {Burgasser}, Adam J. and {Greene}, Jenny E. and {Pan}, Zhiwei and {Shapley}, Alice E. and {Sun}, Fengwu and {Venkatraman}, Padmavathi and {Wang}, Feige},
        title = "{NEXUS: A Search for Nuclear Variability with the First Two JWST NIRCam Epochs}",
      journal = {arXiv e-prints},
         year = 2025,
        month = sep,
          eid = {arXiv:2509.19585},
        pages = {arXiv:2509.19585},
          doi = {10.48550/arXiv.2509.19585},
archivePrefix = {arXiv},
       eprint = {2509.19585},
 primaryClass = {astro-ph.GA},
       adsurl = {https://ui.adsabs.harvard.edu/abs/2025arXiv250919585S}
}

@ARTICLE{Tee+2025,
       author = {{Tee}, Wei Leong and {Fan}, Xiaohui and {Wang}, Feige and {Yang}, Jinyi},
        title = "{Lack of Rest-frame Ultraviolet Variability in Little Red Dots Based on HST and JWST Observations}",
      journal = {\apjl},
         year = 2025,
        month = apr,
       volume = {983},
       number = {1},
          eid = {L26},
        pages = {L26},
          doi = {10.3847/2041-8213/adc5e3},
archivePrefix = {arXiv},
       eprint = {2412.05242},
 primaryClass = {astro-ph.GA},
       adsurl = {https://ui.adsabs.harvard.edu/abs/2025ApJ...983L..26T}
}

@ARTICLE{Burke+2025,
       author = {{Burke}, Colin J. and {Stone}, Zachary and {Shen}, Yue and {Jiang}, Yan-Fei},
        title = "{Too Quiet for Comfort: Local Little Red Dots Lack Variability over Decades}",
      journal = {arXiv e-prints},
         year = 2025,
        month = nov,
          eid = {arXiv:2511.16082},
        pages = {arXiv:2511.16082},
          doi = {10.48550/arXiv.2511.16082},
archivePrefix = {arXiv},
       eprint = {2511.16082},
 primaryClass = {astro-ph.GA},
       adsurl = {https://ui.adsabs.harvard.edu/abs/2025arXiv251116082B}
}

@ARTICLE{Kokubo_Harikane_2025,
       author = {{Kokubo}, Mitsuru and {Harikane}, Yuichi},
        title = "{Challenging the Active Galactic Nucleus Scenario for JWST/NIRSpec Little Red Dot and Non─Little Red Dot Broad H{\ensuremath{\alpha}} Emitters in Light of Nondetection of NIRCam Photometric Variability and X-Ray}",
      journal = {\apj},
         year = 2025,
        month = dec,
       volume = {995},
       number = {1},
          eid = {24},
        pages = {24},
          doi = {10.3847/1538-4357/ae119e},
archivePrefix = {arXiv},
       eprint = {2407.04777},
 primaryClass = {astro-ph.GA},
       adsurl = {https://ui.adsabs.harvard.edu/abs/2025ApJ...995...24K}
}

@ARTICLE{Gloudemans+2025,
       author = {{Gloudemans}, Anniek J. and {Duncan}, Kenneth J. and {Eilers}, Anna-Christina and {Farina}, Emanuele Paolo and {Harikane}, Yuichi and {Inayoshi}, Kohei and {Lambrides}, Erini and {Vardoulaki}, Eleni},
        title = "{Another Piece to the Puzzle: Radio Detection of a JWST-detected Active Galactic Nucleus Candidate}",
      journal = {\apj},
         year = 2025,
        month = jun,
       volume = {986},
       number = {2},
          eid = {130},
        pages = {130},
          doi = {10.3847/1538-4357/adddb9},
archivePrefix = {arXiv},
       eprint = {2501.04912},
 primaryClass = {astro-ph.GA},
       adsurl = {https://ui.adsabs.harvard.edu/abs/2025ApJ...986..130G}
}

@ARTICLE{Tanaka+2025,
       author = {{Tanaka}, Takumi S. and {Akins}, Hollis B. and {Harikane}, Yuichi and {Silverman}, John D. and {Casey}, Caitlin M. and {Inayoshi}, Kohei and {Schindler}, Jan-Torge and {Shimasaku}, Kazuhiro and {Kocevski}, Dale D. and {Onoue}, Masafusa and {Faisst}, Andreas L. and {Robertson}, Brant E. and {Kokorev}, Vasily and {Shuntov}, Marko and {Koekemoer}, Anton M. and {Franco}, Maximilien and {Egami}, Eiichi and {Liu}, Daizhong and {Taylor}, Anthony J. and {Kartaltepe}, Jeyhan S. and {Bosman}, Sarah E.~I. and {Champagne}, Jaclyn B. and {Kakiichi}, Koki and {Harish}, Santosh and {Zhang}, Zijian and {Newman}, Sophie L. and {Kakkad}, Darshan and {Fei}, Qinyue and {Fujimoto}, Seiji and {Li}, Mingyu and {Finkelstein}, Steven L. and {Li}, Zi-Jian and {Lambrides}, Erini and {Sommovigo}, Laura and {Zavala}, Jorge A. and {Ito}, Kei and {Liu}, Zhaoxuan and {Treister}, Ezequiel and {Aravena}, Manuel and {Gozaliasl}, Ghassem and {Zhang}, Haowen and {Hatamnia}, Hossein and {Umeda}, Hiroya and {Inoue}, Akio K. and {Yang}, Jinyi and {Ando}, Makoto and {Arita}, Junya and {Ding}, Xuheng and {Matsui}, Suin and {Shibanuma}, Yuki and {Magdis}, Georgios and {Zhuang}, Mingyang and {Fan}, Xiaohui and {Li}, Zihao and {Liu}, Weizhe and {Lyu}, Jianwei and {Rhodes}, Jason and {Toft}, Sune and {Wang}, Feige and {Zou}, Siwei and {Arango-Toro}, Rafael C. and {Battisti}, Andrew J. and {Gillman}, Steven and {Khostovan}, Ali Ahmad and {Long}, Arianna S. and {Mobasher}, Bahram and {Sanders}, David B.},
        title = "{Discovery of a Little Red Dot Candidate at z {\ensuremath{\gtrsim}} 10 in COSMOS-web Based on MIRI-NIRCam Selection}",
      journal = {\apj},
         year = 2025,
        month = dec,
       volume = {995},
       number = {1},
          eid = {21},
        pages = {21},
          doi = {10.3847/1538-4357/ae145f},
archivePrefix = {arXiv},
       eprint = {2508.00057},
 primaryClass = {astro-ph.GA},
       adsurl = {https://ui.adsabs.harvard.edu/abs/2025ApJ...995...21T}
}

@ARTICLE{Planck+2020,
       author = {{Planck Collaboration} and {Aghanim}, N. and {Akrami}, Y. and {Ashdown}, M. and {Aumont}, J. and {Baccigalupi}, C. and {Ballardini}, M. and {Banday}, A.~J. and {Barreiro}, R.~B. and {Bartolo}, N. and {Basak}, S. and {Battye}, R. and {Benabed}, K. and {Bernard}, J.-P. and {Bersanelli}, M. and {Bielewicz}, P. and {Bock}, J.~J. and {Bond}, J.~R. and {Borrill}, J. and {Bouchet}, F.~R. and {Boulanger}, F. and {Bucher}, M. and {Burigana}, C. and {Butler}, R.~C. and {Calabrese}, E. and {Cardoso}, J.-F. and {Carron}, J. and {Challinor}, A. and {Chiang}, H.~C. and {Chluba}, J. and {Colombo}, L.~P.~L. and {Combet}, C. and {Contreras}, D. and {Crill}, B.~P. and {Cuttaia}, F. and {de Bernardis}, P. and {de Zotti}, G. and {Delabrouille}, J. and {Delouis}, J.-M. and {Di Valentino}, E. and {Diego}, J.~M. and {Dor{\'e}}, O. and {Douspis}, M. and {Ducout}, A. and {Dupac}, X. and {Dusini}, S. and {Efstathiou}, G. and {Elsner}, F. and {En{\ss}lin}, T.~A. and {Eriksen}, H.~K. and {Fantaye}, Y. and {Farhang}, M. and {Fergusson}, J. and {Fernandez-Cobos}, R. and {Finelli}, F. and {Forastieri}, F. and {Frailis}, M. and {Fraisse}, A.~A. and {Franceschi}, E. and {Frolov}, A. and {Galeotta}, S. and {Galli}, S. and {Ganga}, K. and {G{\'e}nova-Santos}, R.~T. and {Gerbino}, M. and {Ghosh}, T. and {Gonz{\'a}lez-Nuevo}, J. and {G{\'o}rski}, K.~M. and {Gratton}, S. and {Gruppuso}, A. and {Gudmundsson}, J.~E. and {Hamann}, J. and {Handley}, W. and {Hansen}, F.~K. and {Herranz}, D. and {Hildebrandt}, S.~R. and {Hivon}, E. and {Huang}, Z. and {Jaffe}, A.~H. and {Jones}, W.~C. and {Karakci}, A. and {Keih{\"a}nen}, E. and {Keskitalo}, R. and {Kiiveri}, K. and {Kim}, J. and {Kisner}, T.~S. and {Knox}, L. and {Krachmalnicoff}, N. and {Kunz}, M. and {Kurki-Suonio}, H. and {Lagache}, G. and {Lamarre}, J.-M. and {Lasenby}, A. and {Lattanzi}, M. and {Lawrence}, C.~R. and {Le Jeune}, M. and {Lemos}, P. and {Lesgourgues}, J. and {Levrier}, F. and {Lewis}, A. and {Liguori}, M. and {Lilje}, P.~B. and {Lilley}, M. and {Lindholm}, V. and {L{\'o}pez-Caniego}, M. and {Lubin}, P.~M. and {Ma}, Y.-Z. and {Mac{\'\i}as-P{\'e}rez}, J.~F. and {Maggio}, G. and {Maino}, D. and {Mandolesi}, N. and {Mangilli}, A. and {Marcos-Caballero}, A. and {Maris}, M. and {Martin}, P.~G. and {Martinelli}, M. and {Mart{\'\i}nez-Gonz{\'a}lez}, E. and {Matarrese}, S. and {Mauri}, N. and {McEwen}, J.~D. and {Meinhold}, P.~R. and {Melchiorri}, A. and {Mennella}, A. and {Migliaccio}, M. and {Millea}, M. and {Mitra}, S. and {Miville-Desch{\^e}nes}, M.-A. and {Molinari}, D. and {Montier}, L. and {Morgante}, G. and {Moss}, A. and {Natoli}, P. and {N{\o}rgaard-Nielsen}, H.~U. and {Pagano}, L. and {Paoletti}, D. and {Partridge}, B. and {Patanchon}, G. and {Peiris}, H.~V. and {Perrotta}, F. and {Pettorino}, V. and {Piacentini}, F. and {Polastri}, L. and {Polenta}, G. and {Puget}, J.-L. and {Rachen}, J.~P. and {Reinecke}, M. and {Remazeilles}, M. and {Renzi}, A. and {Rocha}, G. and {Rosset}, C. and {Roudier}, G. and {Rubi{\~n}o-Mart{\'\i}n}, J.~A. and {Ruiz-Granados}, B. and {Salvati}, L. and {Sandri}, M. and {Savelainen}, M. and {Scott}, D. and {Shellard}, E.~P.~S. and {Sirignano}, C. and {Sirri}, G. and {Spencer}, L.~D. and {Sunyaev}, R. and {Suur-Uski}, A.-S. and {Tauber}, J.~A. and {Tavagnacco}, D. and {Tenti}, M. and {Toffolatti}, L. and {Tomasi}, M. and {Trombetti}, T. and {Valenziano}, L. and {Valiviita}, J. and {Van Tent}, B. and {Vibert}, L. and {Vielva}, P. and {Villa}, F. and {Vittorio}, N. and {Wandelt}, B.~D. and {Wehus}, I.~K. and {White}, M. and {White}, S.~D.~M. and {Zacchei}, A. and {Zonca}, A.},
        title = "{Planck 2018 results. VI. Cosmological parameters}",
      journal = {\aap},
         year = 2020,
        month = sep,
       volume = {641},
          eid = {A6},
        pages = {A6},
          doi = {10.1051/0004-6361/201833910},
archivePrefix = {arXiv},
       eprint = {1807.06209},
 primaryClass = {astro-ph.CO},
       adsurl = {https://ui.adsabs.harvard.edu/abs/2020A&A...641A...6P}
}

@ARTICLE{Ubler+2025,
       author = {{{\"U}bler}, Hannah and {Mazzolari}, Giovanni and {Maiolino}, Roberto and {D'Eugenio}, Francesco and {Davari}, Nazanin and {Juod{\v{z}}balis}, Ignas and {Schneider}, Raffaella and {Valiante}, Rosa and {Arribas}, Santiago and {Bertola}, Elena and {Bunker}, Andrew J. and {Bromm}, Volker and {Carniani}, Stefano and {Charlot}, St{\'e}phane and {Cresci}, Giovanni and {Curti}, Mirko and {Davies}, Richard and {Eisenhauer}, Frank and {Fabian}, Andrew and {F{\"o}rster Schreiber}, Natascha M. and {Genzel}, Reinhard and {Inayoshi}, Kohei and {Ivey}, Lucy R. and {Jones}, Gareth C. and {Liu}, Boyuan and {Lutz}, Dieter and {Mackenzie}, Ruari and {Matthee}, Jorryt and {Parlanti}, Eleonora and {Perna}, Michele and {Robertson}, Brant and {Rodr{\'\i}guez del Pino}, Bruno and {Shimizu}, T. Taro and {Sijacki}, Debora and {Sturm}, Eckhard and {Tacchella}, Sandro and {Tacconi}, Linda and {Tozzi}, Giulia and {Trinca}, Alessandro and {Venturi}, Giacomo and {Volonteri}, Marta and {Willot}, Chris and {Zhang}, Saiyang},
        title = "{BlackTHUNDER: evidence for three massive black holes in a z\raisebox{-0.5ex}\textasciitilde5 galaxy}",
      journal = {arXiv e-prints},
         year = 2025,
        month = sep,
          eid = {arXiv:2509.21575},
        pages = {arXiv:2509.21575},
          doi = {10.48550/arXiv.2509.21575},
archivePrefix = {arXiv},
       eprint = {2509.21575},
 primaryClass = {astro-ph.GA},
       adsurl = {https://ui.adsabs.harvard.edu/abs/2025arXiv250921575U}
}

@ARTICLE{Finkelstein+2022,
       author = {{Finkelstein}, Steven L. and {Bagley}, Micaela B. and {Arrabal Haro}, Pablo and {Dickinson}, Mark and {Ferguson}, Henry C. and {Kartaltepe}, Jeyhan S. and {Papovich}, Casey and {Burgarella}, Denis and {Kocevski}, Dale D. and {Huertas-Company}, Marc and {Iyer}, Kartheik G. and {Koekemoer}, Anton M. and {Larson}, Rebecca L. and {P{\'e}rez-Gonz{\'a}lez}, Pablo G. and {Rose}, Caitlin and {Tacchella}, Sandro and {Wilkins}, Stephen M. and {Chworowsky}, Katherine and {Medrano}, Aubrey and {Morales}, Alexa M. and {Somerville}, Rachel S. and {Yung}, L.~Y. Aaron and {Fontana}, Adriano and {Giavalisco}, Mauro and {Grazian}, Andrea and {Grogin}, Norman A. and {Kewley}, Lisa J. and {Kirkpatrick}, Allison and {Kurczynski}, Peter and {Lotz}, Jennifer M. and {Pentericci}, Laura and {Pirzkal}, Nor and {Ravindranath}, Swara and {Ryan}, Russell E. and {Trump}, Jonathan R. and {Yang}, Guang and {Almaini}, Omar and {Amor{\'\i}n}, Ricardo O. and {Annunziatella}, Marianna and {Backhaus}, Bren E. and {Barro}, Guillermo and {Behroozi}, Peter and {Bell}, Eric F. and {Bhatawdekar}, Rachana and {Bisigello}, Laura and {Bromm}, Volker and {Buat}, V{\'e}ronique and {Buitrago}, Fernando and {Calabr{\`o}}, Antonello and {Casey}, Caitlin M. and {Castellano}, Marco and {Ch{\'a}vez Ortiz}, {\'O}scar A. and {Ciesla}, Laure and {Cleri}, Nikko J. and {Cohen}, Seth H. and {Cole}, Justin W. and {Cooke}, Kevin C. and {Cooper}, M.~C. and {Cooray}, Asantha R. and {Costantin}, Luca and {Cox}, Isabella G. and {Croton}, Darren and {Daddi}, Emanuele and {Dav{\'e}}, Romeel and {de La Vega}, Alexander and {Dekel}, Avishai and {Elbaz}, David and {Estrada-Carpenter}, Vicente and {Faber}, Sandra M. and {Fern{\'a}ndez}, Vital and {Finkelstein}, Keely D. and {Freundlich}, Jonathan and {Fujimoto}, Seiji and {Garc{\'\i}a-Argum{\'a}nez}, {\'A}ngela and {Gardner}, Jonathan P. and {Gawiser}, Eric and {G{\'o}mez-Guijarro}, Carlos and {Guo}, Yuchen and {Hamblin}, Kurt and {Hamilton}, Timothy S. and {Hathi}, Nimish P. and {Holwerda}, Benne W. and {Hirschmann}, Michaela and {Hutchison}, Taylor A. and {Jaskot}, Anne E. and {Jha}, Saurabh W. and {Jogee}, Shardha and {Juneau}, St{\'e}phanie and {Jung}, Intae and {Kassin}, Susan A. and {Le Bail}, Aur{\'e}lien and {Leung}, Gene C.~K. and {Lucas}, Ray A. and {Magnelli}, Benjamin and {Mantha}, Kameswara Bharadwaj and {Matharu}, Jasleen and {McGrath}, Elizabeth J. and {McIntosh}, Daniel H. and {Merlin}, Emiliano and {Mobasher}, Bahram and {Newman}, Jeffrey A. and {Nicholls}, David C. and {Pandya}, Viraj and {Rafelski}, Marc and {Ronayne}, Kaila and {Santini}, Paola and {Seill{\'e}}, Lise-Marie and {Shah}, Ekta A. and {Shen}, Lu and {Simons}, Raymond C. and {Snyder}, Gregory F. and {Stanway}, Elizabeth R. and {Straughn}, Amber N. and {Teplitz}, Harry I. and {Vanderhoof}, Brittany N. and {Vega-Ferrero}, Jes{\'u}s and {Wang}, Weichen and {Weiner}, Benjamin J. and {Willmer}, Christopher N.~A. and {Wuyts}, Stijn and {Zavala}, Jorge A. and {Ceers Team}},
        title = "{A Long Time Ago in a Galaxy Far, Far Away: A Candidate z {\ensuremath{\sim}} 12 Galaxy in Early JWST CEERS Imaging}",
      journal = {\apjl},
         year = 2022,
        month = dec,
       volume = {940},
       number = {2},
          eid = {L55},
        pages = {L55},
          doi = {10.3847/2041-8213/ac966e},
archivePrefix = {arXiv},
       eprint = {2207.12474},
 primaryClass = {astro-ph.GA},
       adsurl = {https://ui.adsabs.harvard.edu/abs/2022ApJ...940L..55F}
}

@ARTICLE{Inayoshi_Ho_2025,
       author = {{Inayoshi}, Kohei and {Ho}, Luis C.},
        title = "{A Critical Evaluation of the Physical Nature of the Little Red Dots}",
      journal = {arXiv e-prints},
         year = 2025,
        month = dec,
          eid = {arXiv:2512.03130},
        pages = {arXiv:2512.03130},
          doi = {10.48550/arXiv.2512.03130},
archivePrefix = {arXiv},
       eprint = {2512.03130},
 primaryClass = {astro-ph.GA},
       adsurl = {https://ui.adsabs.harvard.edu/abs/2025arXiv251203130I}
}

@software{Brammer_2023,
       author = {{Brammer}, Gabriel},
        title = "{grizli}",
         year = 2023,
        month = sep,
          eid = {10.5281/zenodo.8370018},
          doi = {10.5281/zenodo.8370018},
      version = {1.9.11},
    publisher = {Zenodo},
       adsurl = {https://ui.adsabs.harvard.edu/abs/2023zndo...8370018B}
}
\bibliographystyle{aasjournalv7}

\end{document}